\documentclass[10pt,preprint,longabstract]{aastex}

\begin{document}


\title{The Sloan Digital Sky Survey Quasar Catalog~IV. Fifth Data Release
}
\author{
Donald P.\ Schneider\altaffilmark{1}
Patrick B.\ Hall\altaffilmark{2},
Gordon T.\ Richards\altaffilmark{3,4},
Michael A.\ Strauss\altaffilmark{5},
Daniel E.\ Vanden Berk\altaffilmark{1},
Scott F.\ Anderson\altaffilmark{6},
W.\ N.\ Brandt\altaffilmark{1},
Xiaohui Fan\altaffilmark{7},
Sebastian Jester\altaffilmark{8,9},
Jim Gray\altaffilmark{10},
James E.\ Gunn\altaffilmark{5},
Mark U.\ SubbaRao\altaffilmark{11},
Anirudda R.\ Thakar\altaffilmark{3},
Chris Stoughton\altaffilmark{12},
Alexander S.\ Szalay\altaffilmark{3},
Brian Yanny\altaffilmark{12},
Donald G.\ York\altaffilmark{13,14},
Neta A.\ Bahcall\altaffilmark{5},
J.\ Barentine\altaffilmark{15},
Michael R.\ Blanton\altaffilmark{16},
Howard Brewington\altaffilmark{15},
J.\ Brinkmann\altaffilmark{15},
Robert J.\ Brunner\altaffilmark{17},
Francisco J.\ Castander\altaffilmark{18},
Istv\'an Csabai\altaffilmark{19},
Joshua A.\ Frieman\altaffilmark{20,12,13},
Masataka Fukugita\altaffilmark{21},
Michael Harvanek\altaffilmark{15},
David W.\ Hogg\altaffilmark{16},
\v{Z}eljko Ivezi\'{c}\altaffilmark{6},
Stephen M.\ Kent\altaffilmark{12,13},
S.\ J.\ Kleinman\altaffilmark{15},
G.\ R.\ Knapp\altaffilmark{5},
Richard G.\ Kron\altaffilmark{13,12},
Jurek Krzesi\'{n}ski\altaffilmark{22},
Daniel C.\ Long\altaffilmark{15},
Robert H.\ Lupton\altaffilmark{5},
Atsuko Nitta\altaffilmark{23},
Jeffrey R.\ Pier\altaffilmark{24},
David H.\ Saxe\altaffilmark{25},
Yue Shen\altaffilmark{5},
Stephanie A.\ Snedden\altaffilmark{15},
David H.\ Weinberg\altaffilmark{26},
Jian Wu\altaffilmark{1}
}

\altaffiltext{1}{Department of Astronomy and Astrophysics, The Pennsylvania State University, 525 Davey Laboratory, University Park, PA 16802.}
\altaffiltext{2}{
  Department of Physics \& Astronomy,
   York University, 4700 Keele Street, Toronto, Ontario, M3J 1P3, Canada.
}
\altaffiltext{3}{
  Department of Physics and Astronomy,
   The Johns Hopkins University,
   3400 North Charles Street, Baltimore, MD 21218-2686.
}
\altaffiltext{4}{
  Department of Physics,
   Drexel University, 3141 Chestnut Street, Philadelphia, PA 19104.
}
\altaffiltext{5}{
  Princeton University Observatory, Peyton Hall, Princeton, NJ 08544.
}
\altaffiltext{6}{
  Department of Astronomy,
   University of Washington, Box 351580, Seattle, WA 98195.
}
\altaffiltext{7}{
  Steward Observatory,
   University of Arizona, 933 North Cherry Avenue, Tucson, AZ 85721.
}
\altaffiltext{8}{
  School of Physics and Astronomy,
   University of Southampton, Southampton SO17 1BJ, UK.
}
\altaffiltext{9}{
  Max-Planck-Institut f\"{u}r Astronomie,
   K\"{o}nigstuhl 17, D-69117 Heidelberg, Germany.
}
\altaffiltext{10}{
  Microsoft Research, 301 Howard Street, No. 830, San Francisco, CA 94105.
}
\altaffiltext{11}{
  University of Chicago and Adler Planetarium and Astronomy Museum,
   1300 S. Lake Shore Drive, Chicago, IL 60605.
}
\altaffiltext{12}{
  Fermi National Accelerator Laboratory, P.O. Box 500, Batavia, IL 60510.
}
\altaffiltext{13}{
  Department of Astronomy and Astrophysics,
   The University of Chicago, 5640 South Ellis Avenue, Chicago, IL 60637.
}
\altaffiltext{14}{
  Enrico Fermi Institute,
   The University of Chicago, 5640 South Ellis Avenue, Chicago, IL 60637.
}
\altaffiltext{15}{
  Apache Point Observatory, P.O. Box 59, Sunspot, NM 88349.
}
\altaffiltext{16}{
  Department of Physics,
   New York University, 4 Washington Place, New York, NY 10003.
}
\altaffiltext{17}{
  Department of Astronomy,
   University of Illinois at Urbana-Champaign,
   1002 West Green Street, Urbana, IL 61801-3080.
}
\altaffiltext{18}{
  Institut de Ci\`encies de l'Espai (CSIC-IEEC),
   Campus UAB, 08193 Bellaterra, Barcelona, Spain.
}
\altaffiltext{19}{
  Department of Physics of Complex Systems,
   E\"otv\"os University, Budapest, Pf.\ 32, H-1518 Budapest, Hungary.
}
\altaffiltext{20}{
  Center for Cosmological Physics,
   The University of Chicago, 5640 South Ellis Avenue Chicago, IL 60637.
}
\altaffiltext{21}{
  Institute for Cosmic Ray Research,
   University of Tokyo, 5-1-5 Kashiwa, Kashiwa City, Chiba 277-8582, Japan.
}
\altaffiltext{22}{
  Mt. Suhora Observatory,
   Cracow Pedagogical University, ul. Podchorazych 2, 30-084 Cracow, Poland.
}
\altaffiltext{23}{
  Gemini Observatory, 670 North A'ohoku Place, Hilo, HI 96720.
}
\altaffiltext{24}{
  US Naval Observatory,
   Flagstaff Station, P.O. Box 1149, Flagstaff, AZ 86002.
}
\altaffiltext{25}{
  490 Wilson's Crossing Road, Auburn, NH 03032.
}
\altaffiltext{26}{
  Department of Astronomy,
   Ohio State University, 140 West 18th Avenue, Columbus, OH 43210-1173.
}


\begin{abstract}
We present the fourth edition of the Sloan Digital Sky Survey (SDSS)
Quasar Catalog.  The catalog
contains 77,429 objects; this is an increase of over 30,000
entries since the previous edition.  The catalog consists of the objects
in the SDSS Fifth Data Release that have luminosities larger
than \hbox{$M_{i} = -22.0$} (in a cosmology with
\hbox{$H_0$ = 70 km s$^{-1}$ Mpc$^{-1}$,}
\hbox{$\Omega_M$ = 0.3,}
and \hbox{$\Omega_{\Lambda}$ = 0.7),} have at least one
emission line with FWHM larger than 1000~km~s$^{-1}$
or have interesting/complex absorption features,
are fainter than \hbox{$i \approx 15.0$,}
and have highly
reliable redshifts.
The area covered by the catalog is~$\approx$~5740~deg$^2$.
The quasar redshifts range from~0.08 to~5.41, with a median value of~1.48;
the catalog includes 891 quasars at
redshifts greater than four, of which 36 are at redshifts greater than five.
Approximately half of the catalog quasars have \hbox{$i < 19$;}
nearly all have \hbox{$i< 21$.}
For each object the catalog presents positions accurate
to better than~0.2$''$~rms per coordinate,
five-band ($ugriz$) CCD-based photometry with typical accuracy
of~0.03~mag, and information on the morphology and selection method.
The catalog also contains basic radio, near-infrared, and X-ray emission
properties of the quasars, when available, from other large-area surveys.
The calibrated digital spectra cover the wavelength region
3800--9200 \AA\ at
a spectral resolution \hbox{of $\simeq$ 2000;} the spectra can be retrieved
from the public database using the information provided in the catalog.
The average SDSS colors of quasars as a function of redshift, derived from
the catalog entries, are presented in tabular form.
Approximately 96\% of the objects in the catalog were discovered by the SDSS.
\end{abstract}


\section{Introduction}
This paper describes the Fourth Edition of the Sloan Digital Sky Survey
(SDSS; York et al.~2000) Quasar Catalog.  Previous versions of the
catalog (Schneider et al.~2002, 2003, 2005; hereafter Papers~I, II, and~III)
were published
with the SDSS Early Data Release (EDR; Stoughton et al.~2002),
the SDSS First Data Release (DR1; Abazajian et al.~2003), and the
SDSS Third Data Release (DR3; Abazajian et al.~2005), and contained 3,814,
16,713, and 46,420 quasars, respectively.
The current catalog is the entire set of quasars from the SDSS-I
Quasar Survey; the SDSS-I was completed on 30 June~2005 and the Fifth
Data Release (DR5; Adelman-McCarthy et al.~2007) was made public
on 30~June~2006.
The catalog contains
77,429 quasars, the vast majority of which were discovered by the SDSS.
The SDSS Quasar Survey is continuing via the SDSS-II Legacy Survey, which
is is an extension of the SDSS-I.

The catalog in the present paper consists of the DR5 objects that
have a luminosity larger than
\hbox{$M_{i} = -22.0$}  (calculated assuming an
\hbox{$H_0$ = 70 km s$^{-1}$ Mpc$^{-1}$,} \hbox{$\Omega_M$ = 0.3,}
\hbox{$\Omega_{\Lambda}$ = 0.7} cosmology [Spergel et al.~2006],
which will be used throughout this paper), and whose SDSS
spectra contain at least one broad emission line
(velocity FWHM larger than \hbox{$\approx$ 1000 km s$^{-1}$)}
or have interesting/complex absorption-line features.
The catalog also has a bright limit \hbox{of $i \approx 15.0$.}
The quasars range in redshift from~0.08 to~5.41; 78\%~have redshifts
below~2.0.

The objects are denoted in the catalog by their DR5
J2000 coordinates;
the format for the object name
is \hbox{SDSS Jhhmmss.ss+ddmmss.s}.  Since the image data used for the
astrometric information can change between data releases (e.g., a region
with poor seeing that is included in an early release is superseded by
a newer observation in good seeing),
the coordinates for an object can change at
the~0.1$''$ to~0.2$''$ level; hence
the designation of a given source can change between data releases.  Except on
very rare occasions (see \S 5.1), this change in position is much less
than~1$''$.
When merging SDSS Quasar Catalogs with previous databases one should
always use the coordinates, not object names, to identify unique entries.

The DR5 catalog does not include classes of Active Galactic Nuclei (AGN)
such as Type~2 quasars, Seyfert galaxies, and BL~Lacertae objects; studies
of these sources in the SDSS
can be found in Zakamska et al.~(2003) (Type~2), Kauffmann et al.~(2003)
and Hao et al.~(2005) (Seyferts), and
Collinge et al.~(2005) and Anderson et al.~(2007) (BL Lacs).  Spectra of the
highest redshift SDSS quasars \hbox{($z > 5.7$;}
e.g., Fan et al.~2003, 2006a) were not acquired as
part of the SDSS quasar survey (the objects were identified as candidates in
the SDSS imaging data, but the spectra were not obtained with the
SDSS spectrographs),
so they are not included in the catalog.

The observations used to produce the catalog are presented in
Section 2; the construction of the catalog and the catalog format
are discussed in Sections 3 and~4, respectively.  Section~5
presents an overview of the catalog, and a summary is given in Section~6.
The catalog is presented in an electronic table in this paper and
can also be found at an SDSS public web
site.\footnote{\tt
http://www.sdss.org/dr5/products/value$\_$added/qsocat$\_$dr5.html}

\section{Observations}

\subsection{Sloan Digital Sky Survey}

The Sloan Digital Sky Survey
uses a CCD camera \hbox{(Gunn et al. 1998)} on a
dedicated \hbox{2.5-m} telescope
\hbox{(Gunn et al. 2006)}
at Apache Point Observatory,
New Mexico, to obtain images in five broad optical bands ($ugriz$;
Fukugita et al.~1996) over approximately
10,000~deg$^2$ of the high Galactic latitude sky.
The
survey data-processing software measures the properties of each detected object
in the imaging data in all five bands, and determines and applies both
astrometric and photometric
calibrations (Pier et al., 2003; \hbox{Lupton et al. 2001};
Ivezi\'c et al.~2004).
Photometric calibration is provided by simultaneous
observations with a 20-inch telescope at the same site (see Hogg
et al.~2001, Smith et al.~2002, Stoughton et al.~2002, and
Tucker et al.~2006).
The SDSS photometric system is based on the AB magnitude scale
(Oke \& Gunn~1983).

The catalog contains photometry from~204
SDSS imaging runs acquired between
19~September~1998 (Run~94) and 13~May~2005 (Run~5326).

\subsection{Target Selection}

The SDSS filter system was designed to identify quasars at redshifts between
zero and approximately six;
most quasar candidates are selected based on
their location in multidimensional SDSS color-space.
The Point Spread Function (PSF) magnitudes are used for the quasar
target selection, and the selection is based on magnitudes and colors
that have been corrected for Galactic extinction
(using the maps of Schlegel, Finkbeiner, \& Davis~1998).
An $i$ magnitude limit of~19.1
is imposed for candidates whose colors indicate
a probable redshift of less than~$\approx$~3.0 (selected from the $ugri$
color cube);
high-redshift candidates (selected from the $griz$ color cube)
are accepted if \hbox{$i < 20.2$} and the source is unresolved.
The errors on the $i$ measurements
are typically \hbox{0.02--0.03} and \hbox{0.03--0.04} magnitudes at the
brighter and fainter limits, respectively.
In addition to the multicolor selection, unresolved objects brighter
\hbox{than $i = 19.1$} that lie within~2.0$''$ of a FIRST radio source
(Becker, White, \&~Helfand~1995) are also identified as primary quasar
candidates.
Target selection also imposes a maximum brightness limit
\hbox{($i \approx 15.0$)} on quasar candidates; the spectra of objects that
exceed this brightness could contaminate the adjacent spectra
on the detectors of the SDSS spectrographs.
A detailed description of the quasar selection process and possible
biases can be
found in Richards et al.~(2002a).

The primary sample described above was supplemented by
quasars that were targeted by
the following SDSS spectroscopic target selection algorithms:
Galaxy and Luminous Red Galaxy (Strauss et al.~2002 and
Eisenstein et al.~2001),
X-ray (object near the position of a {\it ROSAT} All-Sky Survey
[RASS; Voges et al.~1999,~2000] source; see Anderson et al.~2003),
Star (point source with a color typical of an interesting class of star),
or Serendipity (unusual color or FIRST matches).  The SDSS is designed to
be complete in the Galaxy, Luminous Red Galaxy and Quasar programs,
(in practice various limitations reduce the completeness to about~90\%)
but no attempt at completeness was made for the other categories.
Most of the DR5 quasars that
fall below the magnitude limits of the quasar survey were selected by
the serendipity algorithm (see \S 5).

While the bulk of the catalog objects targeted as quasars were selected
based on the algorithm of Richards et al.~(2002a), during the early years
of the SDSS the quasar selection software was undergoing constant modification
to improve its efficiency.  All of the sources in Papers~I and~II, and some of
the Paper~III objects, were not identified with the final selection
algorithm.
Once the final target selection software was
installed, the algorithm was applied to the entire SDSS photometric database.
Each DR5 quasar
has two spectroscopic target selection flags listed in the catalog:
BEST, which refers to the final algorithm, and TARGET, which is the target
flag used in the actual spectroscopic targeting.  There are also two sets
of photometric measurements for each quasar: BEST, which
refers to the measurements
with the latest photometric software on the highest quality data, and TARGET,
which are the values used at the time of the spectroscopic target selection.

Extreme care must be exercised when constructing statistical samples
from this catalog; if one uses the values produced by only the latest version
of the selection software, not only must one drop the catalog
quasars that were not
identified as quasar candidates by the final selection software, one must also
account for
quasar candidates produced by the final version that were not observed in the
SDSS spectroscopic survey (this can occur
in regions of sky whose spectroscopic targets were
identified by early versions of the selection software).
The selection for the UV-excess quasars,
which comprise the majority ($\approx$ 80\%) of the objects in the
DR5 Catalog, has remained
reasonably uniform; the changes to the selection algorithm were primarily
designed to increase the effectiveness of the identification of
\hbox{$3.0 < z < 3.8$} quasars.
Extensive discussion of the
completeness and efficiency of the selection can be found in
Richards et al.~(2002a) and
Vanden~Berk et al.~(2005);
Richards et al.~(2006) describes the process for 
the construction of
statistical SDSS quasar samples (see also Adelman-McCarthy et al.~2007).
The survey efficiency (the ratio of
quasars to quasar candidates) for the ultraviolet excess-selected
candidates, which comprise
the bulk of the quasar sample, is about~77\%.  (The catalog contains
information
on which objects can be used in a uniform sample; see Section~4.)

\subsection{Spectroscopy}

Spectroscopic targets chosen by the various SDSS selection algorithms
(i.e., quasars, galaxies, stars, serendipity) are arranged onto
a series of 3$^{\circ}$ diameter circular fields (Blanton et al.~2003).
Details of the spectroscopic observations can be found in
York et al.~(2000), Castander et al.~(2001), Stoughton et al.~(2002),
and Paper~I.
A total of~1458 spectroscopic fields, taken between 5~March~2000 and
14~June~2005, provided the quasars for the DR5 quasar catalog;
the locations of the plate centers
can be found from the information given by Adelman-McCarthy et al.~(2007).
The DR5 spectroscopic program attempted to cover, in a well-defined manner,
an area of~$\approx$~5740~deg$^2$.  Spectroscopic plate~716 was the first
spectroscopic observation that was based on the
final version of the quasar target
selection algorithm of Richards et al.~(2002a);
the detailed tiling
information in the SDSS database must be consulted to identify those regions
of sky targeted with the final selection algorithm (see Richards et al.~2006).

The two SDSS double spectrographs produce data covering \hbox{3800--9200 \AA }
at a spectral resolution \hbox{of $\simeq$ 2000.}
The data, along with the associated calibration frames, are processed by
the SDSS spectroscopic pipeline (see Stoughton et al.~2002).
The calibrated spectra are classified into various groups
(e.g., star, galaxy, quasar), and redshifts are determined by two independent
software packages.
Objects whose spectra cannot be classified
by the software are flagged for visual inspection.
Figure~1 shows the calibrated SDSS spectra of four previously unknown
catalog quasars representing a range of properties.
The processed DR5 spectra {\it have not} been corrected for Galactic
extinction.

\section{Construction of the SDSS DR5 Quasar Catalog}

The quasars in the catalog were drawn from three sets of SDSS observations:
1)~the primary survey area, 2)~``Bonus" plates, which are spectroscopic
observations of regions near to, but outside of, the primary survey area,
and 3)~``Special" plates, where the spectroscopic targets were not chosen
by the standard SDSS target selection algorithms (e.g., a set of plates
to investigate the structure of the Galaxy; see Adelman-McCarthy et al.~2006).

The DR5 quasar catalog was constructed, as were the previous editions,
in three stages: 1)~Creation of a
quasar candidate database, 2)~Visual examination of the spectra of the
quasar candidates, and 3)~Application of luminosity
and emission-line velocity width criteria.  All three tasks were initially
done without reference to the material in the previous SDSS Quasar Catalogs,
although the results of each task were compared to the Paper~III database
(e.g., the construction of the quasar database was not viewed as complete
until it was understood why any Paper~III quasars were not included).

\subsection{Creation of the DR5 Quasar Candidate Database}

This catalog of bona-fide quasars, that have redshifts checked
by eye and luminosities and line widths that meet the formal quasar
definition, is constructed from a larger ``master" table of confirmed
quasars and quasar candidates.  This master table was
created using an SQL query to the public SDSS-DR5 database
(i.e., the Catalog Archive Server [CAS];
http://cas.sdss.org/astrodr5/).  Two versions of the photometric
database exist, which contain the properties of objects when targeted
for spectroscopic observations
(TARGET) and as determined in the latest processing (BEST).
These databases are divided
into multiple tables and subtables to facilitate access
to only the most relevant data for a particular use.  In the case of
the quasar catalog construction, we have made use of the {\tt PhotoObjAll}
and {\tt SpecObjAll} tables, which contain, respectively, the photometric
information for {\em all} SDSS sources and for {\em all} SDSS spectra.
In the case of {\tt PhotoObjAll}, both the TARGET and BEST versions are
queried.  These tables include duplicate observations of objects and
observations of objects that lie outside of the formal SDSS area
(as compared to the {\tt PhotoObj} and
{\tt SpecObj} tables, which include only sources in the formal SDSS area),
and are the most complete database files.  For example, in
{\tt PhotoObjAll}, two (or more) observations of
a single object may exist; if so, one is
classified as {\tt PRIMARY}, the other(s) as {\tt SECONDARY}.

This master table contains all objects identified as quasar candidate
targets for spectroscopy
in either the TARGET or BEST photometric databases.  Quasar candidates are
those objects
which have had one or more of the following flags set by the algorithm
described by Richards et al.~(2002a):

\noindent
{\tt TARGET\_QSO\_HIZ OR TARGET\_QSO\_CAP OR
TARGET\_QSO\_SKIRT OR TARGET\_QSO\_FIRST\_CAP\\
 OR TARGET\_QSO\_FIRST\_SKIRT}

\noindent
\hbox{( = 0x0000001F,} except for the ``special" plates [see
Adelman-McCarthy et al.~2006, 2007], where additional care is
required in interpreting the flags).

Objects flagged as {\tt TARGET\_QSO\_MAG\_OUTLIER}
and {\tt TARGET\_QSO\_REJECT} are
not included, as these flags are meant only for diagnostic purposes.
(In the CAS documentation and the EDR paper,
{\tt TARGET\_QSO\_MAG\_OUTLIER} is called {\tt TARGET\_QSO\_FAINT}.)
Furthermore, the master table includes any objects with spectra that have
been classified by the spectroscopic pipeline
as quasars ({\tt specClass=QSO} or {\tt HIZ\_QSO}),
that have {\tt UNKNOWN} type, or that have
redshifts greater than~0.6.  (On
rare occasions the spectroscopic pipeline
measures the correct redshift for a quasar
but classifies the object as a galaxy.)

The query was run on the union of the
database tables {\tt Target..PhotoObjAll}, {\tt Best..SpecObjAll},
and {\tt Best..PhotoObjAll}.
Multiple entries for a given object are retained at this stage.

Ten objects in the DR3 Quasar Catalog were missed by this query.
One omission was due to an ``unmapped" fiber (a spectrum of a quasar was
obtained, but because of a failure in the mapping of fiber number to
location in the sky, we are no longer certain of the celestial position of the
object); the other nine were low-redshift
AGN that were not classified as quasars by the spectroscopic pipeline
(this result provides an estimate of the incompleteness produced by the query).
We were able to identify the information for all ten quasars in the database
and add the material to the master table.

Four automated cuts were made to the master table database of
329,884 candidates
\footnote{The master table is known as the QSOConcordanceALL table, which
can be found in the SDSS database; see \hbox{
\tt http://cas.sdss.org/astrodr5/en/help/browser/description.asp?n=QsoConcordanceAll\&t=U.}}:
1)~Objects targeted as quasars but whose
spectra had not yet been obtained by the closing date of DR5 (124,447 objects),
2)~Candidates classified with high confidence as ``stars" by the spectroscopic
pipeline that had redshifts less
than~0.002 (33,653), 3)~Objects whose photometric measurements
have not been loaded into the CAS
(3106) and 4)~Multiple spectra (coordinate agreement better than~1.0$''$)
of the same object (40,007).
In cases of duplicate spectra of an object,
the ``science primary" spectrum is selected (i.e., the spectrum was obtained as
part of normal science operations);
when there is more than one science primary observation (or when none of
the spectra have this flag set), the spectrum with the
highest signal-to-noise ratio (S/N) is retained (see Stoughton et al.~2002
for a description of the
science primary flag).  These actions produced a list of~128,671
unique quasar candidates.

\subsection{Visual Examination of the Spectra}

The SDSS spectra of the remaining quasar
candidates were manually inspected by
several of the authors (DPS, PBH, GTR, MAS, and SFA); as in
previous papers in this series, we found
that the
spectroscopic pipeline redshifts and classifications
of the overwhelming majority of the objects
are accurate.
Tens of thousands of objects were dropped from the
list because they were obviously not quasars (these objects tended to be
low S/N stars, unusual stars, and a mix of absorption-line and
narrow emission-line objects); this large number of candidates that
are not quasars is due to the
inclusive nature of our initial database query.
Spectra for which redshifts could not be determined (low signal-to-noise
ratio or subject to data-processing difficulties) were also removed from
the sample.
This visual inspection resulted in the revisions of
the redshifts of 863 quasars; the changes in the individual redshifts were
usually quite substantial, due to the spectroscopic pipeline
misidentifying emission lines.

An independent determination of the redshifts \hbox{of 5,865}
quasars with redshifts larger than~2.9
in the catalog was performed by Shen et al.~(2007).  The redshift
differences between the two sets of measurements
follow a Gaussian distribution (with slightly
extended wings), with a mean
of~0.002 and a dispersion of~0.01.

The catalog contains numerous examples of
extreme Broad Absorption Line (BAL) Quasars
(see Hall et al.~2002); it is difficult
if not impossible to apply the emission-line width criterion for these objects,
but they are clearly of interest, have more in common with ``typical" quasars
than with narrow-emission line galaxies,
and have historically been included in quasar
catalogs.  We have included in the catalog all objects with broad
absorption-line spectra that meet the \hbox{$M_i < -22.0$}
luminosity criterion.

\subsection{Luminosity and Line Width Criteria}

As in Papers~II and~III, we adopt a luminosity limit of
\hbox{$M_{i} = -22.0$.}
The absolute magnitudes were calculated by correcting the BEST $i$
measurement for Galactic extinction (using the maps of
Schlegel, Finkbeiner, \& Davis~1998) and assuming that the quasar
spectral energy distribution in the ultraviolet-optical
can be represented by a power law
\hbox{($f_{\nu} \propto \nu^{\alpha}$),} where $\alpha$~=~$-0.5$
(Vanden~Berk et al.~2001).  (In the 134 cases where BEST photometry
was not available, the TARGET measurements were substituted for the
absolute magnitude calculation.)
This approach ignores the contributions
of emission lines and the observed distribution in continuum slopes.
Emission lines can contribute several tenths of a magnitude to the
k-correction (see Richards et al. 2006), and variations in the continuum
slopes can introduce a magnitude or more of error into the calculation
of the absolute magnitude, depending upon the
redshift.
The absolute magnitudes
will be particularly uncertain at redshifts near and above
five, when the Lyman~$\alpha$ emission line (with a typical observed
equivalent width \hbox{of $\approx 400-500$ \AA })
and strong Lyman~$\alpha$ forest absorption enter
the~$i$ bandpass.

Quasars near the $M_i$~=~$-22.0$ luminosity
limit are often not enormously brighter in the $i$-band than the
starlight produced by the host galaxy.  Although the PSF-based
SDSS photometry presented
in the catalog are less susceptible to host galaxy contamination than
are fixed-aperture measurements, the nucleus of the host
galaxy can still
contribute appreciably to this measurement for the lowest luminosity
entries in the catalog (see Hao et al.~2005).
An object of $M_i = -22.0$ will reach the \hbox{$i = 19.1$} ``low-redshift"
selection limit at a redshift of~$\approx$~0.4.

After visual inspection and application of the luminosity criterion had
reduced the number of quasar candidates to under 80,000 objects, the
remaining spectra were processed with an automated line-measuring routine.  The
spectra for objects whose maximum line width was less than 1000~km~s$^{-1}$
were visually examined; if the measurement was deemed to be an accurate
reflection of the line (automated routines occasionally have spectacular
failures when dealing with complex line profiles), the object was removed
from the catalog.

\section{Catalog Format}

The DR5 SDSS Quasar Catalog is available in three types of files at the
SDSS public web site listed in the introduction:
1)~a standard ASCII file with fixed-size columns,
2)~a gzipped compressed version of the ASCII file (which is smaller than
the uncompressed version
by a factor of more than four), and 3)~a binary FITS table format.
The following description applies to the standard ASCII file.  All files
contain the same number of columns, but the storage of the numbers differs
slightly in the ASCII and FITS formats; the FITS header contains all of the
required documentation.  Table~1 provides a summary of the information
contained in each of the columns in the ASCII catalog.

The standard ASCII catalog (Table~2 of this paper)
contains information on~77,429 quasars in
a 36~MB file.
The DR5 format is similar to that of DR3 with a few minor differences.

The first~80 lines consist of catalog documentation; this is followed
by~77,429 lines containing
information on the quasars.  There are~74 columns in each line; a summary
of the information is given in Table~1 (the documentation in the ASCII catalog
header
is essentially an expansion of Table~1).  At least one space separates all the
column entries, and, except for the first and last columns (SDSS designation
and the object name if previously known),
all entries are reported in either floating point or integer format.

Notes on the catalog columns:

\noindent
1) The DR5 object designation, given by the format
\hbox{SDSS Jhhmmss.ss+ddmmss.s}; only the final~18
characters  are listed in the catalog
(i.e., the \hbox{``SDSS J"} for each entry is dropped).
The coordinates in the object name follow
IAU convention and are truncated, not rounded.

\noindent
2--3) The J2000 coordinates (Right Ascension and
Declination) in decimal degrees.  The positions for the vast majority of
the objects are accurate to~0.1$''$~rms or better
in each coordinate; the largest
expected errors are~0.2$''$ (see Pier et al~2003).  The SDSS coordinates
are placed in the International Celestial Reference System, primarily
through the United States Naval Observatory
CCD Astrograph Catalog (Zacharias et al.~2000), and
have an rms accuracy of~0.045$''$ per coordinate.

\noindent
4) The quasar redshifts.
A total of 863 of the CAS redshifts were revised during our visual inspection.
A detailed description of the redshift measurements is given in Section 4.10
of Stoughton et al.~(2002).  A comparison of 299 quasars observed
at multiple epochs by the SDSS (Wilhite et al.~2005) found an rms
difference of~0.006 in the measured redshifts for a given object.  It is
well known that the redshifts of individual broad emission lines
in quasars exhibit significant offsets
from their systemic redshifts (e.g., Gaskell 1982, Richards et al.~2002b,
Shen et al.~2007);
the catalog redshifts attempt to correct for this effect in the ensemble
average (see Stoughton et al.~2002).

%

\noindent
5--14) The DR5 PSF magnitudes and errors (not corrected for Galactic
extinction) from BEST photometry for each object in the five SDSS
filters.  Some of the relevant imaging scans, such as special scans
through M31 (see the DR4 and DR5 papers) were never loaded into the
CAS, therefore the BEST photometry is not available for them.  Thus there
are 134 quasars which have entries of ``0.000" for their BEST
photometric measurements.

  The effective 
wavelengths of the $u$, $g$, $r$, $i$, and $z$ bandpasses are 3541,
4653, 6147, 7461, and~8904~\AA, respectively (for an \hbox{$\alpha =
-0.5$} power-law spectral energy distribution using the definition of
effective wavelength given in Schneider, Gunn, \& Hoessel~1983).  The
photometric measurements are reported in the natural system of the
SDSS camera, and the magnitudes are normalized to the AB system (Oke
\& Gunn~1983).  The measurements are reported as asinh magnitudes
(Lupton, Gunn, \& Szalay~1999); see Adelman-McCarthy et al.~(2007) for
additional discussion and references for the accuracy of the
photometric measurements.  The TARGET PSF photometric measurements are
presented in columns \hbox{63--72.}

\smallskip\smallskip
\vbox{\noindent
15) The Galactic extinction in the $u$ band based on the maps of
Schlegel, Finkbeiner, \& Davis~(1998).  For an $R_V = 3.1$ absorbing medium,
the extinctions in the SDSS bands can be expressed as

$$ A_x \ = \ C_x \ E(B-V) $$

\noindent
where $x$ is the filter ($ugriz$), and values of $C_x$ are
5.155, 3.793, 2.751, 2.086, and 1.479 for $ugriz$, respectively
($A_g$, $A_r$, $A_i$, and $A_z$ are 0.736, 0.534, 0.405, and 0.287 times
$A_u$).
}

\noindent
16) The logarithm of the Galactic neutral hydrogen column density along the
line of sight to the quasar. These values were
estimated via interpolation of the 21-cm data from Stark et al.~(1992),
using the COLDEN software provided by the {\it Chandra} X-ray Center.
Errors associated with the interpolation are typically expected to
be less than $\approx 1\times 10^{20}$~cm$^{-2}$ (e.g., see \S5 of
Elvis, Lockman, \& Fassnacht 1994).

\noindent
17) Radio properties.  If there is a source
in the FIRST catalog
within~2.0$''$ of
the quasar position, this column contains the FIRST
peak flux density at 20~cm encoded as an AB magnitude

$$ AB \ = \ -2.5 \log \left( {f_{\nu} \over 3631 \ {\rm Jy}} \right) $$

\noindent
(see Ivezi\'c et al.~2002).
An entry of ``0.000" indicates no match to a FIRST source; an entry of
``$-1.000$" indicates that the object does not lie in the region covered by
the final catalog of the FIRST survey.  The catalog contains 6226 FIRST
matches; 5729 DR5 quasars lie outside of the FIRST area.

\noindent
18) The S/N of the FIRST source whose flux is given in column~17.

\noindent
19) Separation between the SDSS and FIRST coordinates (in arc seconds).

\noindent
20) In cases when the FIRST counterpart to
an SDSS source is extended, the FIRST catalog position of the source
may differ by more than 2$''$ from the optical position. 
A~``1" in column~20
indicates that no matching FIRST source was found within
2$''$ of the optical position, but that there {\it is}
significant detection (larger than~3$\sigma$)
of FIRST flux at the optical position. This is
the case for 2440 SDSS quasars.

\noindent
21) A ``1" in column~21 identifies
the 1596 sources with a FIRST match in either columns~17 or~20 that also
have at least one FIRST counterpart located between 2.0$''$ (the SDSS-FIRST
matching radius) and
30$''$ of the optical position.
Based on the average FIRST
source surface density of 90~deg$^{-2}$, we \hbox{expect 50--60} of these
matches to be chance superpositions.

\noindent
22) The logarithm
of the vignetting-corrected count rate (photons s$^{-1}$)
in the broad energy band \hbox{(0.1--2.4 keV)} in the
{\it ROSAT} All-Sky Survey Faint Source Catalog (Voges et al.~2000) and the
{\it ROSAT} All-Sky Survey Bright Source Catalog (Voges et al.~1999).
The matching radius was set to~30$''$;
an entry of~``$-9.000$" in this column indicates no X-ray detection.
There are 4133 RASS matches in the DR5 catalog.

\noindent
23) The S/N of the {\it ROSAT} measurement.

\noindent
24) Separation between the SDSS and {\it ROSAT} All-Sky Survey
coordinates (in arc seconds).

\noindent
25--30) The $JHK$ magnitudes and errors from the Two Micron All Sky Survey
(2MASS; Skrutskie et al. 2006)
All-Sky Data Release Point Source Catalog (Cutri et al.~2003) using
a matching radius
of~2.0$''$.  A non-detection by 2MASS is indicated by a ``0.000" in these
columns.  Note that the 2MASS measurements are Vega-based, not AB,
magnitudes.  The catalog contains 9824 2MASS matches.

\noindent
31) Separation between the SDSS and 2MASS coordinates (in arc seconds).

\noindent
32) The absolute magnitude in the $i$ band calculated by correcting for
Galactic extinction and assuming
\hbox{$H_0$ = 70 km s$^{-1}$ Mpc$^{-1}$,}
$\Omega_M$~=~0.3, $\Omega_{\Lambda}$~=~0.7, and a power-law (frequency)
continuum index of~$-0.5$.

\noindent
33) The $\Delta (g-i)$ color, which is the difference in the Galactic
extinction corrected $(g-i)$ for the quasar and that of the mean of the
quasars at that redshift.  If $\Delta (g-i)$ is not defined for the quasar,
which occurs for objects at either \hbox{$z < 0.12$} or \hbox{$z > 5.12$}
the column will contain~``$-9.000$".  See Section~5.2 for a description of this
quantity.

\noindent
34) Morphological information.
If the SDSS photometric pipeline classified the image of the quasar
as a point source, the catalog entry is~0; if the quasar is extended, the
catalog entry is~1.

\noindent
35) The SDSS {\tt SCIENCEPRIMARY} flag, which
indicates whether the spectrum was taken as a normal science spectrum
({\tt SCIENCEPRIMARY}~=~1) or for another purpose
({\tt SCIENCEPRIMARY}~=~0).  The latter category contains
Quality Assurance and calibration spectra, or spectra of objects
located outside of the nominal survey area.  Over 90\%
of the DR5 entries (69,762 objects)
are {\tt SCIENCEPRIMARY}~=~1.

\noindent
36) This flag provides information on
whether the photometric object is designated {\tt PRIMARY}~(1),
{\tt SECONDARY}~(2), or {\tt FAMILY} (3; these are blended objects that have
not been deblended).
During target selection,
only {\tt PRIMARY} objects
are considered (except on occasion for objects located in fields that
are not part of the nominal survey area);
however, differences between TARGET and BEST
photometric pipeline versions make it possible that the BEST
photometric object belonging to a spectrum is either not detected at
all, or is a non-primary object (see \S 3.1 above).  Over~99\% of the
catalog entries are {\tt PRIMARY};
613 quasars are {\tt SECONDARY} and~9 are {\tt FAMILY}.
There are 124 quasars with
an entry of ``0" in this column; each of these is an object that lacks
BEST photometry.
For statistical analysis, one should use only
{\tt PRIMARY} objects; {\tt SECONDARY} and
{\tt FAMILY} objects are included in the catalog for the sake of completeness
with respect to confirmed quasars.

\noindent
37) The ``uniform selection" flag, either 0 or 1; a ``1" indicates
that the object was
identified as a primary quasar target (37,574 catalog entries)
with the final target selection
algorithm as given by Richards et al.~(2002a).  These objects constitute a
statistical sample.

\noindent
38) The 32-bit SDSS target-selection flag from BEST processing
({\tt PRIMTARGET}; see Table~26 in Stoughton et al.~2002 for details); this
is the flag produced by running the selection algorithm of
Richards et al.~(2002a) on the most recent processing of the image data.
The target-selection flag from TARGET processing is found in column~55.

\noindent
39--45) The spectroscopic target selection breakdown (BEST) for each object.
The target selection flag in column~38 is decoded for seven groups:
Low-redshift quasar, High-redshift quasar, FIRST, ROSAT, Serendipity,
Star, and Galaxy
An entry of~``1" indicates that the object satisfied the given criterion
(see Stoughton et al.~2002).  Note that an object can
be, and often is, targeted by more than one selection algorithm.
The last two columns in Table~3 presents the number of quasars
identified by the individual BEST target selection algorithm; the column
labeled ``Sole" indicates the number of objects that were detected by only
one of the seven listed selection algorithms.

\noindent
46--47) The SDSS Imaging Run number and the Modified Julian Date (MJD) of the
photometric observation used in the catalog.  The MJD is given as an integer;
all observations on a given night have the same integer MJD
(and, because of the observatory's location, the same UT date). For example,
imaging run 94 has an MJD of 51075; this observation was taken on
1998 September~19~(UT).

\noindent
48--50) Information about the spectroscopic observation (Modified Julian
Date, spectroscopic plate number, and spectroscopic fiber number) used to
determine the redshift.
These three numbers are unique for each spectrum, and
can be used to retrieve the digital spectra from the public SDSS database.

\noindent
51--54) Additional SDSS processing information: the
photometric processing rerun number; the camera column (1--6) containing
the image of the object, the field number of the run containing the object,
and the object identification number
(see Stoughton et al.~2002 for descriptions of these parameters).

\noindent
55) The 32-bit SDSS target selection flag from the TARGET processing, i.e.,
the value that was used when the spectroscopic plate was drilled.  This
may not match the BEST target selection flag because a different versions of
the selection algorithm were used, the selection was done with different
image data (superior quality data of the field was obtained after the
spectroscopic observations were completed), or
different processings of the same data were used.  Objects with no TARGET flag
were either identified as quasars as a result of Quality Assurance observations
and/or from special plates with somewhat different targeting criteria
(see Adelman-McCarthy~2006).

\noindent
56--62) The spectroscopic target-selection breakdown (TARGET) for each object;
this is the same convention as followed in columns~39--45 for the BEST
target-selection flag.

\noindent
63--72) The DR5 PSF
magnitudes and errors (not corrected for Galactic reddening) from TARGET
photometry.

\noindent
73) The 64-bit integer that uniquely describes the spectroscopic observation
that is listed in the catalog (SpecObjID).

\noindent
74) Name of object in the NASA/IPAC Extragalactic Database (NED).
If there is a source in the NED quasar database within~5.0$''$ of the
quasar position, the NED object name is given in this column.
The NED quasar database contains over 100,000
objects.  Occasionally
NED will list the SDSS name for objects that were not discovered by the SDSS.

\section{Catalog Summary}

The 77,429 objects in the catalog represent an increase of 31,009 quasars
over the Paper~III database; of the entries in the new catalog,
74,297 (96.0\%) were discovered by the SDSS (with the caveat that NED
is not complete).
The catalog quasars span a wide range of properties: redshifts
from~0.078 to~5.414, \hbox{$ 14.94 < i < 22.36$}
(506~objects \hbox{have $i > 20.5$;} only~26
have \hbox{$i > 21.0$}),
and \hbox{$ -30.27 < M_{i} < -22.00$.}
The catalog contains 6226, 4133, and~9824
matches to the FIRST, RASS, and 2MASS catalogs, respectively.
The RASS and 2MASS catalogs cover essentially all of the DR5 area, but~5729
(7\%) of
the entries in the DR5 catalog lie outside of the FIRST region.

Figure~2 displays the distribution of the DR5 quasars in the $i$-redshift plane
(the 26 objects with \hbox{$i > 21$} are not plotted).
Objects which NED indicates were previously discovered by investigations other
than the SDSS
are indicated with open circles.  The curved cutoff on the left
hand side of the graph is produced by the minimum luminosity criterion
\hbox{($M_i < -22.0$).}  The ridge in the contours at
\hbox{$i \approx 19.1$}
for redshifts below three reflects the flux limit of the
low-redshift sample; essentially all of the
large number of \hbox{$z < 3$} points with \hbox{$i > 19.1$}
are quasars selected via criteria other than the primary
multicolor sample.

A histogram of the catalog redshifts is shown in the upper curve in
Figure~3.  A clear
majority of
quasars have redshifts below two (the median redshift is~1.48, the
mode is~$\approx$~1.85),
but there is a significant tail
of objects extending out to redshifts beyond five
\hbox{($z_{\rm max}$ = 5.41).}  The dips in the curve at redshifts
of~2.7 and~3.5 arise because the SDSS colors of quasars at these redshifts
are similar to the colors of stars; we decided to accept significant
incompleteness at these redshifts rather than be overwhelmed by a large number
of stellar contaminants in the spectroscopic survey.  Improvements in the
quasar target selection algorithm since the initial editions of the
SDSS Quasar Catalog have increased the efficiency of target
selection at redshifts near~3.5 (compare Figure~3 with Paper~II's
Figure~4; see Richards et al.~2002a for a discussion of the incompleteness
of the SDSS Quasar Survey).

This structure in the catalog redshift histogram can be understood by careful
modelling of the selection effects (e.g., accounting for emission line
effects and using only objects selected in regions whose spectroscopic
observations were chosen with the final version of the quasar target selection
algorithm; also see Figure~8 in Richards et al.~2006).
Repeating the analysis of Richards et al.~(2006) for the
DR5 sample reveals no
structure in the redshift distribution after selection effects have
been included (see lower histogram in Figure~3); this is in contrast to
the reported redshift structure found in the SDSS quasar survey by Bell \&
McDiarmid~(2006).  To construct the lower histogram we have partially removed
the effect of host galaxy contamination (by excluding extended
objects), limited the sample to a uniform magnitude limit of $i<19.1$
(accounting for emission-line effects), and have corrected for
the known incompleteness near $z\sim2.7$ and $z\sim3.5$ due to quasar
colors lying close to or in the stellar locus.  Accounting for selection
effects significantly reduces the number of objects as compared with the raw,
more heterogeneous catalog, but the smaller, more homogeneous sample is what
should be used for statistical analyses.

The distribution of the observed $i$ magnitude
(not corrected for Galactic extinction) of the quasars is given in Figure~4.
The sharp drops in the histogram at \hbox{$i \approx 19.1$} and
\hbox{$i \approx 20.2$} are due to the magnitude limits in the low and
high-redshift samples, respectively.

Figures~5 and~6 display the distribution of the absolute~$i$ magnitudes of the
catalog quasars.  There
is a roughly symmetric peak centered at \hbox{$M_i = -26$} with a FWHM
of approximately one magnitude.  The histogram declines sharply at
high luminosities (only~1.5\% of the objects have \hbox{$M_i < -28.0$)}
and has a gradual decline toward lower luminosities, partially due to
host-galaxy contribution.

A summary of the spectroscopic selection, for both the TARGET and the BEST
algorithms, is given in Table~3.  We report
seven selection classes in the catalog (columns~39 to~45 for BEST,
\hbox{56--62} for TARGET).  Each selection version has two columns,
the number of objects that
satisfied a given selection criterion and
the number of objects that were identified only by that selection
class.  About two-thirds of the catalog entries
were selected based on the SDSS
quasar selection criteria (either a low-redshift or high-redshift candidate,
or both).  Slightly more than
half of the quasars
in the catalog are serendipity-flagged candidates,
which is also primarily an ``unusual
color" algorithm; about one-fifth of the catalog was selected by
the serendipity criteria alone.

Of the~50,093 DR5 quasars that have Galactic-absorption corrected TARGET
$i$ magnitudes brighter than~19.1,
48,593 (97.0\%) were identified by the TARGET quasar multicolor
selection; if one combines TARGET multicolor and FIRST selection (the primary
quasar target selection criteria),
all but~1015 of the \hbox{$i < 19.1$} objects were selected.  (The spectra
of many of the last category of objects were obtained in observations that
were not part of the primary survey.)  The numbers
are similar if one uses the BEST photometry and selection, although the
completeness is not quite as high as with TARGET values.

\subsection{Discrepancies Between the DR5 and Other Quasar Catalogs}

The DR3 database is entirely contained in that of DR5, but there are
66 quasars from Paper~III (out of 46,420 objects) that do not have a
counterpart within~1.0$''$ of a DR5 quasar.  Three of these
``missing" quasars are in the DR5 list; changes in celestial position of
1.1$''$, 1.8$''$, and~5.3$''$ between DR3 and DR5 caused these
quasars to be missed with the~1.0$''$ matching criterion.  The other 63 cases
(0.14\% of the DR3 total) were individually investigated.  Three DR3
objects were dropped because the latest photometry reduced their
luminosities below the catalog limit.  The remaining 60
objects were removed
because 1)~the visual examination of the spectrum either
convinced us that
the object was not a quasar or that the S/N was insufficient to assign
a redshift with confidence or 2)~The widest line in the
latest fit to the spectrum had a FWHM of less than
1000~km~s$^{-1}$.  It should be noted that there have been no changes to
the DR3 spectra in the DR5 database; the missing objects reflect the
inherent uncertainties involved with interpreting objects that either lie
near survey cutoffs or have spectra of marginal~S/N.

There are 40 and 136 DR5 quasars that have redshifts that differ by more
than~0.1 from the DR3 and NED values, respectively (there is, of course,
considerable overlap in these two groups).
In all cases the DR5 measurements are
considered more reliable than those presented in previous publications.
The 40 objects with \hbox{$\vert z_{\rm DR5} - z_{\rm DR3} \vert > 0.1$}
are listed in Table~4.

\subsection{Quasar Colors}

It has long been known that the majority of quasars inhabit a restricted range
in photometric color, and the large sample size and
accurate photometry of the SDSS revealed a relatively tight color-redshift
correlation for quasars (Richards et al.~2001).  This SDSS color relation,
recently presented in Hopkins et al.~(2004), has led to considerable
success in assigning photometric redshifts to quasars (e.g., Weinstein
et al.~2004 and references therein).  All photometric measurements used
in these analyses have been corrected for Galactic extinction.

The dependence of the four standard SDSS colors on redshift for the
DR5 quasars is given in Figure~7.  The dashed line in each panel is the
modal relation for the DR5 quasars; the modal relations are tabulated in
Table~5, along with the values for~$(g-i)$.  The figures show an impressively
tight correlation of color with redshift, although the scatter dramatically
increases when the Lyman~$\alpha$ forest dominates the bluer of the passbands
used to form the color.  The distribution near the modal curve is roughly
symmetric, but there is clearly a significant population of ``red" quasars
that has no ``blue" counterpart.

This table is an improvement over previous work in that it is based on
a larger sample size (a factor of four increase since this relation was
last published) and provides
higher redshift resolution (0.01, except near the extrema).
As in Hopkins et al.~(2004), we
compute the mode, rather than the mean or median, as the most representative
quantity.
However, a formal computation of the mode requires binning the data
both in redshift and by color within redshift bins;  therefore we estimated
the mode from the mean and the median.  Typically, the mode is estimated
as \hbox{(3 $\times$ median$-$2 $\times$ mean),} but we found empirically that
\hbox{(2 $\times$ median$-$mean)} appeared to work
better for this sample in terms of tracing the modal ``ridgeline" with
redshift.

For each of the DR5 quasars we provide the quantity $\Delta (g-i)$, which
is defined by

$$ \Delta (g-i) \ =
\ (g-i)_{\rm QSO} \ - \ \langle (g-i) \rangle_{\rm redshift} $$

\noindent
where $\langle(g-i) \rangle_{\rm redshift}$ is
the entry in Table~5 for the redshift of
the quasar.  This ``differential color" provides an estimate of the continuum
properties of the quasar (values above zero indicate that the object has
a redder continuum than the typical quasar at that redshift).

\subsection{Bright Quasars}

Although the spectroscopic survey is limited to objects fainter than
\hbox{$i \approx 15$}, the SDSS continues to discover a number of
``PG-class"
(Schmidt \& Green~1983) objects.  The DR5 catalog contains 81 entries
with \hbox{$i < 16.0$}; 14 of the quasars are not in the NED database or
attributed to the SDSS by
NED.  The spectrum of the brightest post-DR3 discovery,
\hbox{SDSS J165551.37+214601.8} \hbox{($i = 15.62$}, \hbox{$z = 0.15$)},
is presented in Figure~1.  Three of the SDSS-discovered
objects in this catalog have been
added since Paper~III.

\subsection{Luminous Quasars}

There are 103 catalog quasars with \hbox{$M_i < -29.0$}
(3C~273 has \hbox{$M_{i} \approx -26.6$} in our adopted cosmology); 61 were
discovered by the SDSS, and 18 are
published here for the first time.  The redshifts of these quasars lie
between~1.3 and~5.0.
The most luminous quasar in the catalog is
\hbox{2MASSI J0745217+473436} \hbox{(= SDSS J074521.78+473436.2),} at
\hbox{$M_i = -30.27$} \hbox{and $z = 3.22$.}
Spectra of the two most luminous
post-DR3 discoveries, with absolute $i$ magnitudes of $-29.94$ and $-29.65$,
are displayed in the upper two panels of Figure~1.  The spectra of both
quasars possess a considerable number of absorption features redward of
the Lyman~$\alpha$ emission line.

\subsection{Broad Absorption Line Quasars}

The SDSS quasar selection algorithm has proven to be effective
at finding a wide variety of Broad Absorption Line (BAL) Quasars.  An EDR
sample of 118 BAL quasars was presented by Tolea, Krolik, \& Tsvetanov~(2002).
There have been two editions of the SDSS BAL Quasar Catalog; the first,
associated with Paper~I, contained 224 BAL quasars (Reichard et al.~2003);
the second was based on the
Paper~III catalog and presents 4787 BAL quasars (Trump et al.~2006).
BAL quasars are usually recognized by the presence of
C~IV absorption features, which are only visible in SDSS spectra at
$z > 1.6$, thus the frequency of the BAL quasar phenomenon cannot be found from
simply taking the ratio of BAL quasars to total number of quasars in the SDSS
catalog.
The SDSS has discovered a wide variety of extreme BAL quasars
(see Hall et al.~2002); the lower right panel in Figure~1 presents the
spectrum of an unusual FeLoBAL quasar with strong Balmer absorption
(see Hall~2007 for a discussion of this object).

\subsection{Quasars with Redshifts Below 0.15}

The catalog contains~109 quasars with redshifts below~0.15.
All of these objects are
of low luminosity \hbox{($M_i > -24.0$, only three have $M_i < -23.5$)}
because of the \hbox{$i \approx 15.0$}
limit for the spectroscopic sample.  About three-quarters
of these quasars (83) are extended in the SDSS image data.   A total of~40
of the \hbox{$z < 0.15$} quasars were found by the SDSS; 21 have been added
since Paper~III.

\subsection{High-Redshift ($z \ge 4$) Quasars}

At first light of the SDSS, the most distant known quasar was PC~1247+3406
at redshift of 4.897 (Schneider, Schmidt, and Gunn 1991), which had
been discovered seven years earlier.  Within a year of operation, the
SDSS had discovered quasars with redshifts above five (Fan et al. 1999, 2000);
the DR5 catalog contains 60 objects with redshifts greater than that of
PC~1247+3406.

\noindent
In recent years the SDSS has identified
quasars out to a redshift of~6.4 (Fan et al.~2003, 2006b).
Quasars with redshifts larger than~$\approx$~5.7, however,
cannot be found by the SDSS spectroscopic survey because
at these redshifts the observed wavelength of the
Lyman~$\alpha$ emission line is redward of
the $i$ band; at this point quasars become single-filter ($z$) detections.
At the typical $z$-band flux levels for redshift six quasars, there are simply
too many ``false-positives" to undertake automated targeting.
The largest redshift in the DR5 catalog is
\hbox{SDSS J023137.65$-$072854.5} \hbox{at $z = 5.41$}, which was originally
described by Anderson et al.~(2001).

The DR5 catalog contains 891 quasars with redshifts larger than four;
36 entries have redshifts above five (11 above $z=5.2$), which is more than
a factor of two increase since Paper~III.  The spectra of the 20 highest
redshift post-DR3 objects
(all with redshifts greater than
or equal to~4.99) are displayed in Figure~8.  These redshift five spectra
display a striking variety of emission line properties, and include
an impressive BAL \hbox{at $z = 5.27$.}

We have used archival data from {\it Chandra\/}, {\it ROSAT\/}, and
{\it XMM-Newton\/} to check for new \hbox{X-ray} detections of $z>4$ quasars
with unusual emission-line or absorption-line properties; we do
not report all $z>4$ \hbox{X-ray} detections here as there are now more
than 110 already published.\footnote{See
http://www.astro.psu.edu/users/niel/papers/highz-xray-detected.dat for a
list of X-ray detections and references.}
We found three remarkable $z>4$ \hbox{X-ray} detections:
the $z~=~4.26$ BAL quasar SDSS~J133529.45+410125.9,
the $z~=~4.11$ BAL quasar SDSS~J142305.04+240507.8, and
the $z~=~4.50$ quasar SDSS~J150730.63+553710.8, which shows
remarkably strong C~{\sc iv} emission.
None of these objects has sufficient counts for detailed \hbox{X-ray}
spectral analysis, but we have computed their point-to-point
spectral slopes between rest-frame 2500~\AA\ and 2~keV ($\alpha_{\rm ox}$),
adopting the assumptions in \S2 of Brandt et~al. (2002).
SDSS~J133529.45+410125.9 and SDSS~J142305.04+240507.8 were
serendipitously detected in archival {\it Chandra\/} ACIS observations
and have $\alpha_{\rm ox}=-2.19$ and $\alpha_{\rm ox}=-1.52$,
respectively. Comparing these values to the established relation
between $\alpha_{\rm ox}$ and 2500~\AA\ luminosity (e.g.,
Steffen et~al. 2006), we find that SDSS~J133529.45+410125.9 is
notably \hbox{X-ray} weak, indicating likely \hbox{X-ray} absorption as is
often seen in BAL quasars (e.g., Gallagher et~al. 2006) including those
at $z>4$ (Vignali et~al. 2005). In contrast, the
level of \hbox{X-ray} emission from SDSS~J142305.04+240507.8 is consistent
with that from normal, non-BAL quasars; its relatively narrow
UV absorption, for a BAL quasar, may indicate a relatively
small column density of obscuring material. SDSS~J150730.63+553710.8
is weakly detected in a {\it ROSAT\/} PSPC observation and has
$\alpha_{\rm ox}=-1.47$; this level of \hbox{X-ray} emission is nominal
for a quasar of its luminosity.

We have also checked all quasars with $z>5$ for new \hbox{X-ray} detections
and found none; 21 quasars with $z>5$ have previously reported \hbox{X-ray}
detections.

\subsection{Close Pairs}

The mechanical constraint
that SDSS spectroscopic fibers must be separated by~55$''$ on a given plate
makes it difficult for the spectroscopic survey to confirm close pairs
of quasars.  In regions that
are covered by more than one plate, however, it is possible
to obtain spectra of both components of a close pair;
there are~346 pairs of quasars in the catalog with angular separation less than
an arcminute (34 pairs with separations less than 20$''$).
Most of the pairs are chance superpositions, but there
are many sets whose components have similar redshifts,
suggesting that the quasars may be physically
associated.  The typical uncertainty in the measured value of the redshift
difference between two quasars is~0.02; the catalog contains 18 quasar pairs
with separations of less than an arcminute and \hbox{with $\Delta z < 0.02$.}
These pairs, which are excellent candidates for binary quasars, are listed
in Table~6.  Hennawi et al.~(2006) identified over 200 quasar pairs
in the SDSS,
primarily through spectroscopic observations of SDSS
quasar candidates (based on photometric measurements) near known SDSS quasars;
statistical arguments based on a correlation-function analysis suggests
that most of these pairs are indeed physically associated.

\subsection{Morphology}

The images of~3498 of the DR5 quasars are classified as extended by the
SDSS photometric pipeline;~3291~(94\%) have redshifts below one
(there are nine resolved \hbox{$z > 3.0$} quasars).
The majority of the large-redshift ``resolved" quasars are probably measurement
errors, but this sample may also contain a mix of chance superpositions
of quasars and foreground objects or possibly some
small angle separation gravitational lenses (indeed, several lenses
are present in the resolved quasar sample; see Paper~II and Oguri et al.~2006).

\subsection{Matches with Non-optical Catalogs}

A total of 6226 catalog objects are FIRST sources (defined by a SDSS-FIRST
positional offset of less than~2.0$''$).  Note that 226 of the objects
were selected (with TARGET) solely because
they were FIRST matches (all unresolved SDSS
sources brighter than $i=19.1$ that lie within~2.0$''$ of a FIRST source
are targeted by the quasar spectroscopic
selection algorithm).
Extended radio sources may be missed by this matching.  The upper left
panel in Figure~9 contains a histogram of the angular offsets between the
SDSS and FIRST positions; the solid line is the expected distribution assuming
a~0.21$''$ $1\sigma$ Gaussian error in
the relative SDSS/FIRST positions (found by fitting the points with a
separation less than~1.0$''$.  The small-angle separations are well-fit to
the Rayleigh distribution,
but outside of about~0.5$''$ there is an obvious
excess of observed separations.
The number of chance superpositions was
estimated by shifting the quasar positions by~{$\pm 200''$} in declination
and matching
the new coordinates to the FIRST catalog;
only about~0.1\% of the reported FIRST matches are false.
The large ``tail" of this distribution is not likely to be due to
measurement errors but probably arises from extended radio emission that
may not be precisely centered on the optical image.
To recover radio quasars that have offsets of more than~2.0$''$,
we separately identify all objects with a greater than
3$\sigma$ detection of FIRST flux at the optical position (2440 sources).
For these objects as well as those with a FIRST catalog match within
2$''$, we perform a second FIRST catalog search with 30$''$ matching radius
to identify possible radio lobes associated with the quasar, finding
such matches for 1596 sources.

Matches with the {\it ROSAT} All-Sky Survey Bright and
Faint Source Catalogs
were made with a maximum allowed positional offset
of~30$''$; this is the positional coincidence required for the SDSS
{\it ROSAT} target selection code.
The DR5 catalog contains~4133 RASS matches; approximately~1.3\% are expected
to be false identifications based on an analysis similar to that described
in the previous paragraph.  The SDSS-RASS offsets for the DR5 sample are
presented in the upper right panel of Figure~9; the solid curve, which is
the predicted distribution for a $1\sigma$
positional error of~11.1$''$ (fit using all of the points), matches the data
quite well.

$JHK$ photometric measurements for 9824 DR5 quasars were found by using a
matching radius of~2.0$''$ in the
2MASS All-Sky Data Release Point Source Catalog.  No
infrared information was used to select the SDSS spectroscopic targets.
The positional offset histogram, given in the lower left panel of Figure~9,
is considerably tighter than that for the FIRST matches, although the
Rayleigh fit to the separations less than~1.0$''$ is virtually identical
to the FIRST distribution (1$\sigma$ of 0.21$''$).
There are very few 2MASS
identifications with offsets between~1$''$ and~2$''$; virtually all of the
infrared matches are correct.

%
%

\section{Summary}

The lower right panel in Figure~9 charts the progress of the SDSS Quasar
Survey, denoted by the number of spectroscopically-confirmed quasars,
over the duration of SDSS-I.  Although SDSS-I has now been completed,
the SDSS Quasar Survey is continuing under the SDSS-II project.  By
necessity the SDSS spectroscopy lags the SDSS imaging; at the
conclusion of SDSS-I more than 2000 square degrees of SDSS image
data in the Northern Galactic Cap lacked spectroscopic coverage
(Adelman-McCarthy et al.~2007).  A future edition of the SDSS Quasar Catalog
will incorporate the observations from SDSS-II and should contain approximately
100,000 quasars.

\acknowledgments

The publication of this catalog marks the completion of the SDSS-I
Quasar Survey, and we dedicate this work to the memory of
\hbox{John N. Bahcall.}
John was the initial co-chair of the SDSS Quasar Working Group, a
position he held for nearly a decade.  He played a key role in the
formation of the SDSS Collaboration and the design of the SDSS Quasar Survey,
and was a mentor to many of the members of the Quasar Working Group.
We would like to thank Todd Boroson for suggesting several redshift
adjustments to some of the DR3 Quasar Catalog redshifts.
This work was supported in part by National Science Foundation grants
AST-0307582 and AST-0607634 (DPS, DVB, JW), AST-0307384~(XF), and
AST-0307409~(MAS), and by
NASA LTSA grant NAG5-13035 (WNB, DPS).
PBH acknowledges support by NSERC, and
GTR was supported in part by a Gordon and Betty Moore Fellowship in Data
Intensive Sciences at JHU.
XF acknowledges support from an \hbox{Alfred P. Sloan} Fellowship and
a David and Lucile Packard Fellowship in Science and Engineering.
SJ was supported by the Max-Planck-Gesellschaft (MPI f\"ur Astronomie) through
an Otto Hahn fellowship.
CS was supported by the U.S. Department of Energy under contract
\hbox{DE-AC02-76CH03000.}

Funding for the SDSS and SDSS-II has been
provided by the Alfred P. Sloan Foundation,
the Participating Institutions,
the National Science Foundation,
the U.S. Department of Energy,
the National Aeronautics and Space Administration,
the Japanese Monbukagakusho,
the Max Planck Society,
and the Higher Education Funding Council for England.
The SDSS Web site \hbox{is {\tt http://www.sdss.org/}.}

The SDSS is managed by the Astrophysical Research Consortium
(ARC) for the Participating Institutions.  The Participating
Institutions are
the American Museum of Natural History,
Astrophysical Institute of Potsdam,
University of Basel,
Cambridge University,
Case Western Reserve University,
University of Chicago,
Drexel University,
Fermilab,
the Institute for Advanced Study,
the Japan Participation Group,
Johns Hopkins University,
the Joint Institute for Nuclear Astrophysics,
the Kavli Institute for Particle Astrophysics and Cosmology,
the Korean Scientist Group,
the Chinese Academy of Sciences (LAMOST),
Los Alamos National Laboratory,
the Max-Planck-Institute for Astronomy (MPIA),
the Max-Planck-Institute for Astrophysics (MPA),
New Mexico State University,
Ohio State University,
University of Pittsburgh,
University of Portsmouth,
Princeton University,
the United States Naval Observatory,
and the University of Washington.

This research has made use of 1)~the NASA/IPAC Extragalactic Database (NED)
which is operated by the Jet Propulsion Laboratory, California Institute
of Technology, under contract with the National Aeronautics and Space
Administration, and 2)~data products from the Two Micron All Sky
Survey, which is a joint project of the University of
Massachusetts and the Infrared Processing and Analysis Center/California
Institute of Technology, funded by the National Aeronautics
and Space Administration and the National Science Foundation.


\newpage

\begin{figure}
\includegraphics[angle=90,scale=0.70,viewport= 0 0 500 700]{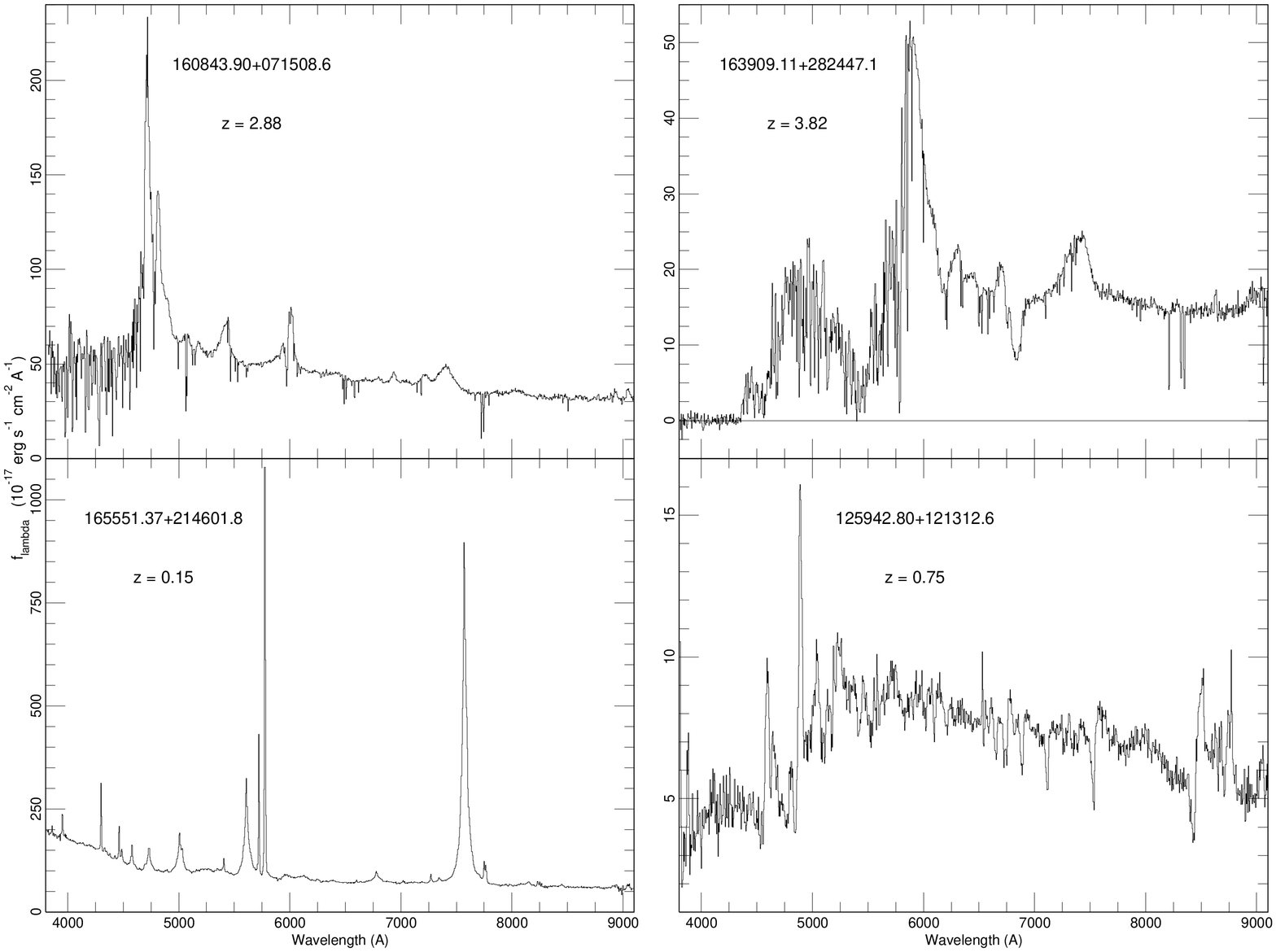}
\caption{
SDSS spectra of four previously unreported quasars.
The spectral
resolution of the data ranges from 1800 to 2100; a dichroic splits the beam
at~6150~\AA .  The data have been rebinned \hbox{to 5 \AA\ pixel$^{-1}$}
for display purposes.
The upper two panels display the two most luminous of the newly
discovered quasars; both objects have \hbox{$M_i < -29.6$.}
\hbox{SDSS J165551.37+214601.8} is the brightest \hbox{($i = 15.62$)} of the
new quasars;
\hbox{SDSS J125942.80+121312.6} is an unusual FeLoBAL quasar with Balmer-line
absorption.
\label{Figure 1 }
}
\end{figure}

\begin{figure}
\includegraphics[angle=0,scale=0.90]{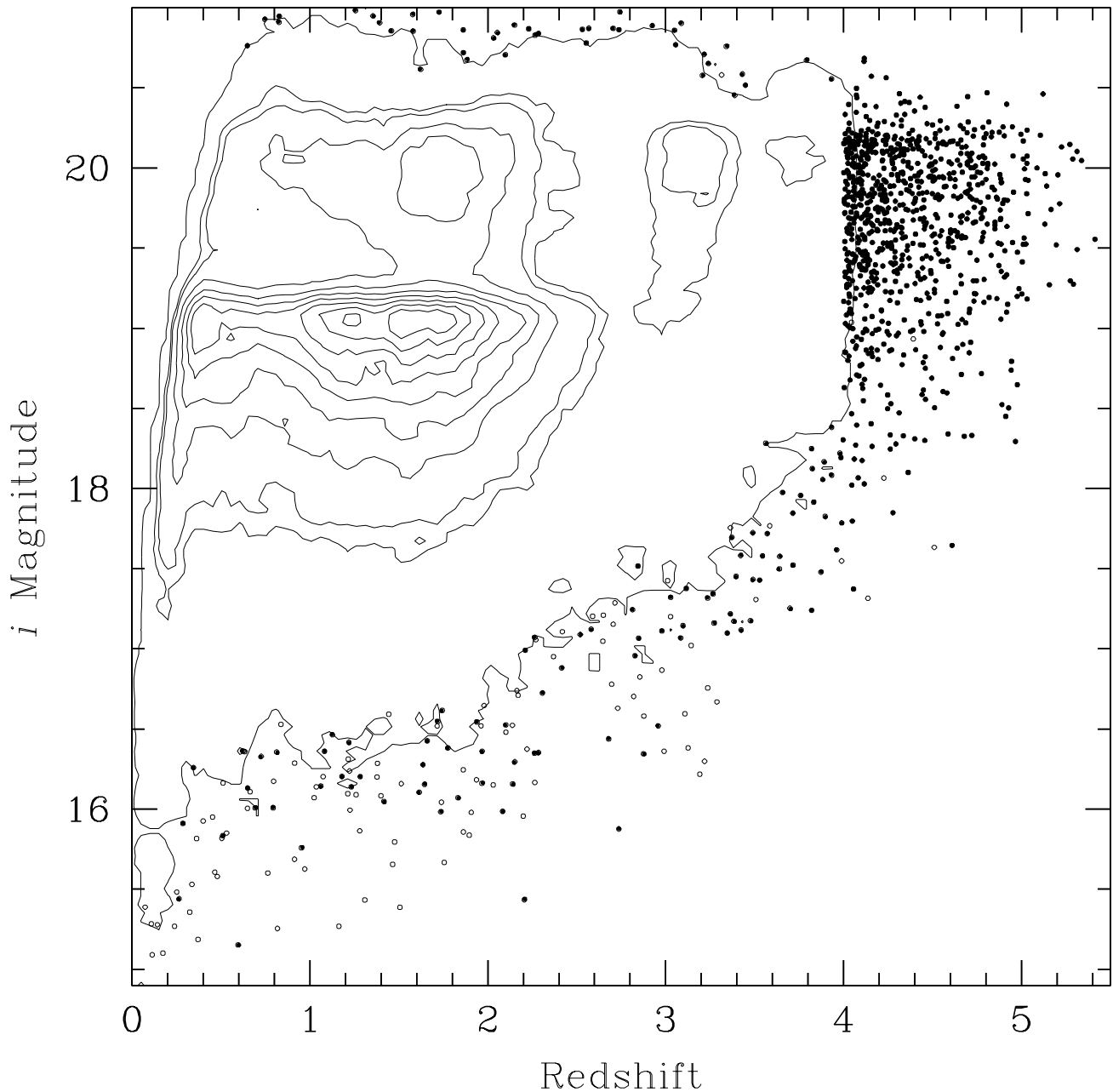}
\figcaption{
The observed~$i$ magnitude as a function of redshift for the~77,429
objects in the catalog.  Open circles indicate quasars in NED that
were recovered but not discovered by
the SDSS.  The 26 quasars with \hbox{$i > 21$} are not plotted.
The distribution is represented by a set of linear contours when the
density of points in this two-dimensional space causes the points to overlap.
The steep gradient at \hbox{$i \approx 19$} is due to the flux limit for the
targeted low-redshift part of the survey;
the dip in the counts at $z$~$\approx$~2.7 arises
because of the high incompleteness of the SDSS Quasar Survey at redshifts
between 2.5 and~3.0 (also see Figure~3).
\label{Figure 2 }
}
\end{figure}

\begin{figure}
\includegraphics[angle=0,scale=0.90]{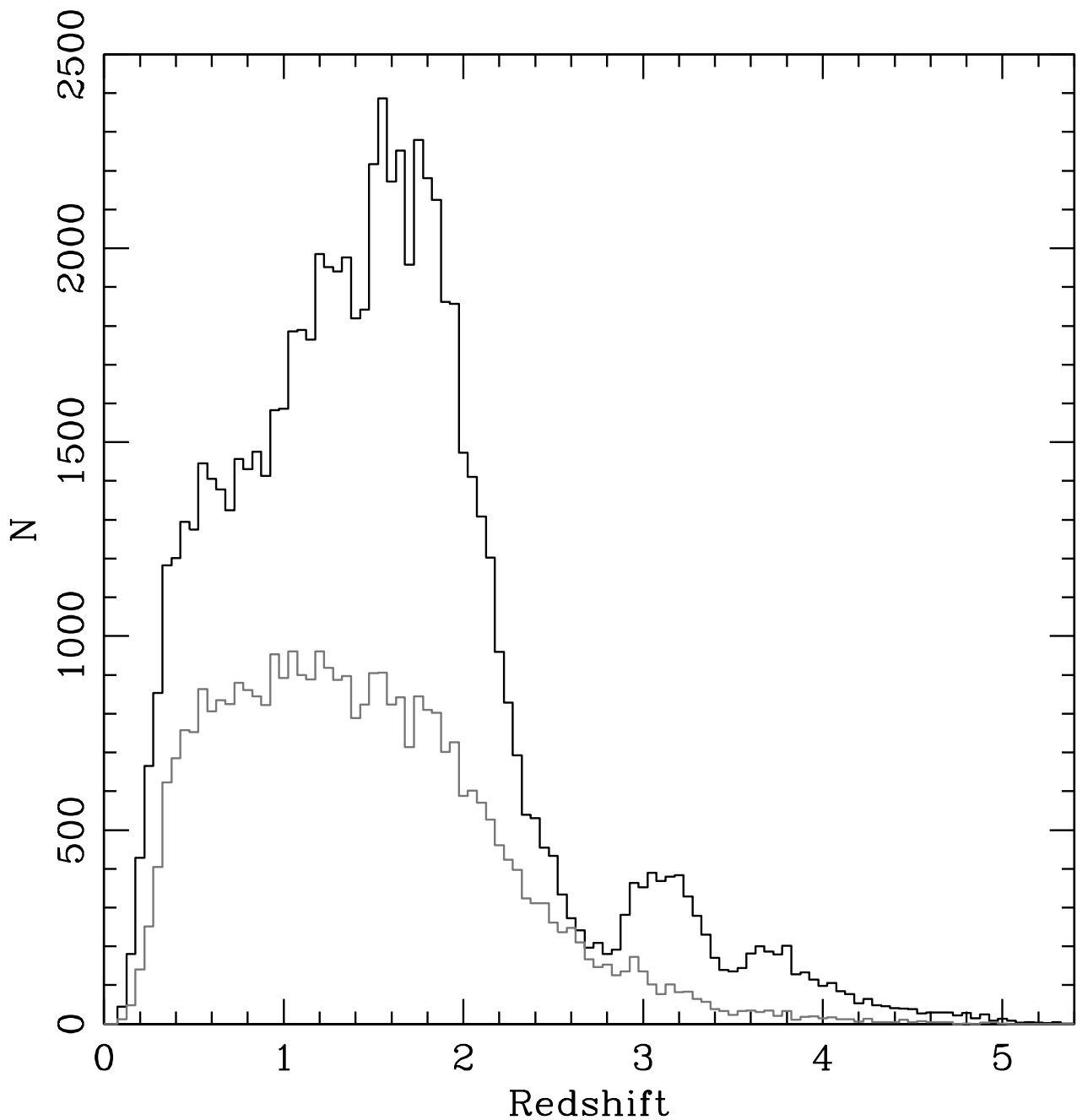}
\figcaption{
The redshift histogram of the catalog quasars.  The redshifts range from~0.08
to~5.41; the median redshift of the catalog is~1.48.
The redshift bins have a width of~0.05.  The dips at redshifts of~2.7 and~3.5
are caused by the reduced efficiency of the selection algorithm at these
redshifts.  The lower histogram is the redshift distribution of
the $i<19.1$ sample after correction
for selection effects (see Section~5).
\label{Figure 3 }
}
\end{figure}

\begin{figure}
\includegraphics[angle=90,scale=1.10,viewport= 0 0 350 700]{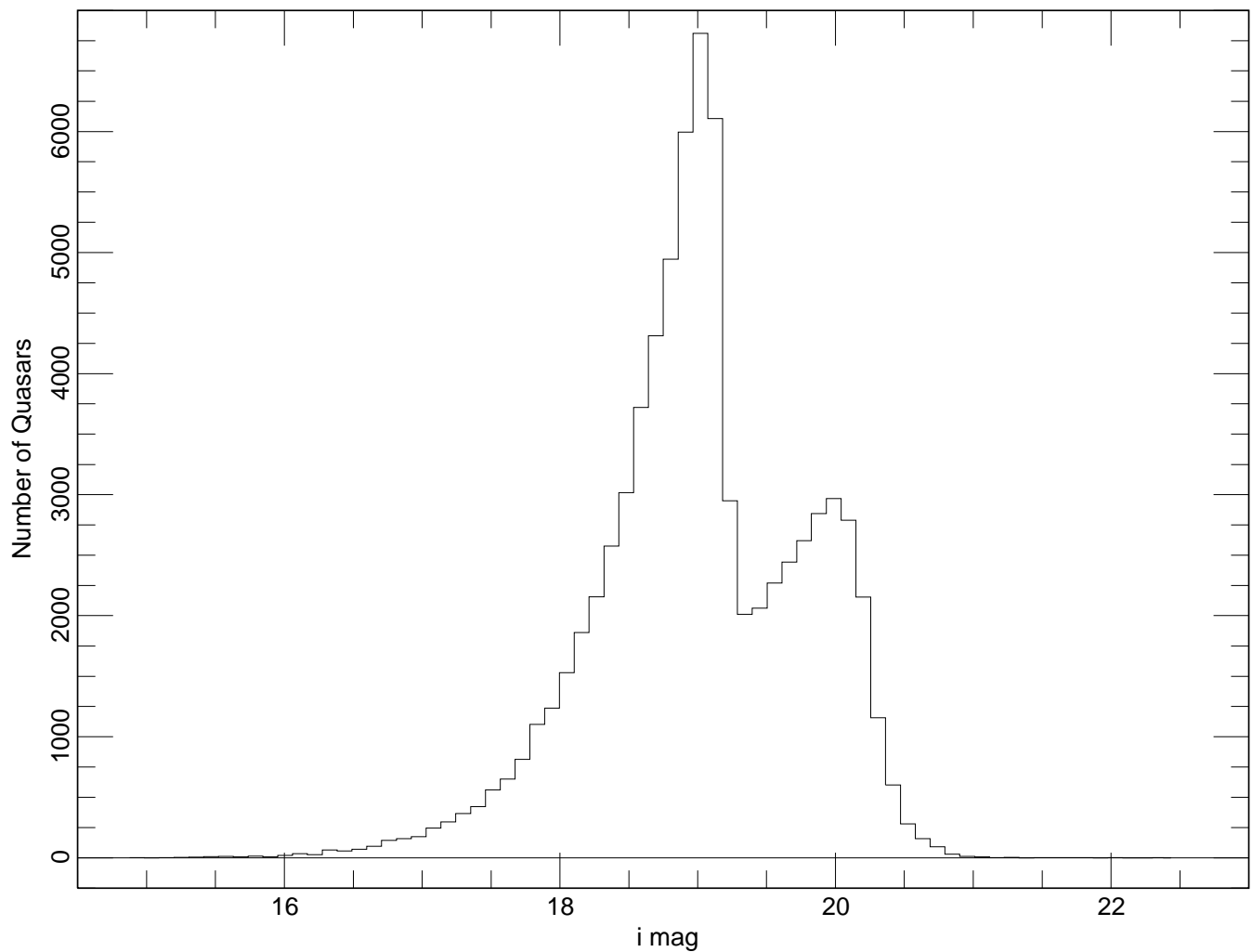}
\figcaption{
The $i$ magnitude (not corrected for Galactic absorption) histogram of the
77,429 catalog quasars.  The magnitude bins have a width of~0.108.  The sharp
drop that occurs at magnitudes slightly fainter than~19 is due to the
flux limit for the low-redshift targeted part of the survey.
Quasars fainter than the
\hbox{$i = 20.2$} high-redshift selection limit were found via other
selection algorithms, primarily serendipity.  The SDSS Quasar survey has
a bright limit of \hbox{$i \approx 15.0$} imposed by the need to avoid
saturation in the spectroscopic observations.
\label{Figure 4 }
}
\end{figure}

\begin{figure}
\includegraphics[angle=0,scale=0.90]{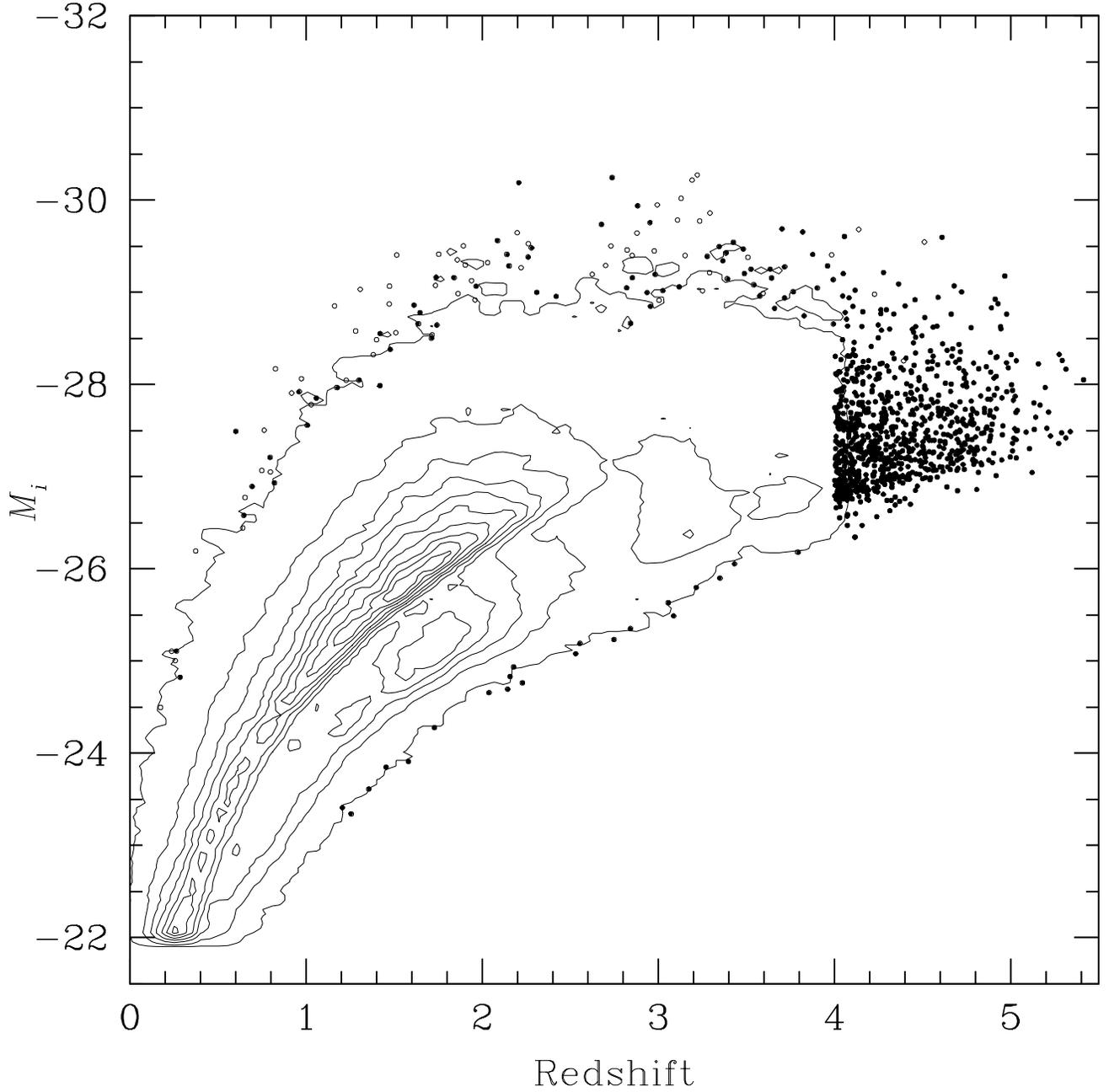}
\figcaption{
The absolute~$i$ magnitude as a function of redshift for the~77,429
objects in the catalog.  Open circles indicate quasars in NED that
were recovered but not discovered by
the SDSS.
The distribution is represented by a set of linear contours when the
density of points in this two-dimensional space causes the points to overlap.
The steep gradient that runs through the midst of the quasar distribution
is produced by the \hbox{$i \approx 19$} flux limit for the
targeted low-redshift part of the survey.
\label{Figure 5 }
}
\end{figure}

\begin{figure}
\includegraphics[angle=90,scale=1.10,viewport= 0 0 350 700]{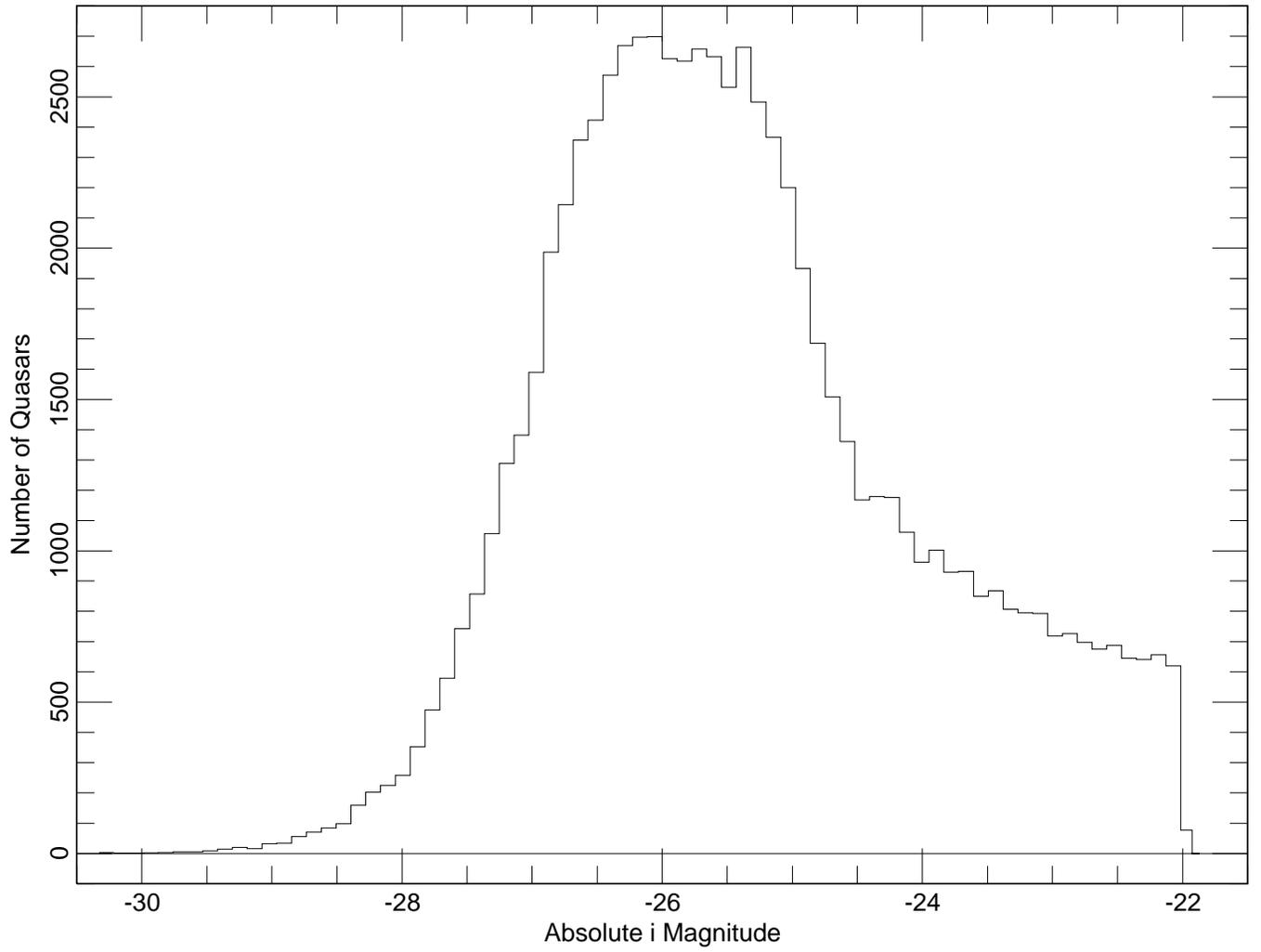}
\figcaption{
The luminosity distribution of the catalog quasars.
The absolute magnitude bins have a width of~0.114.  The most luminous
quasar in the catalog has \hbox{$M_i \approx -30.3$.}  In the adopted cosmology
3C~273 has \hbox{$M_i \approx -26.6$.}
\label{Figure 6 }
}
\end{figure}

\begin{figure}
\includegraphics[angle=0,scale=0.90]{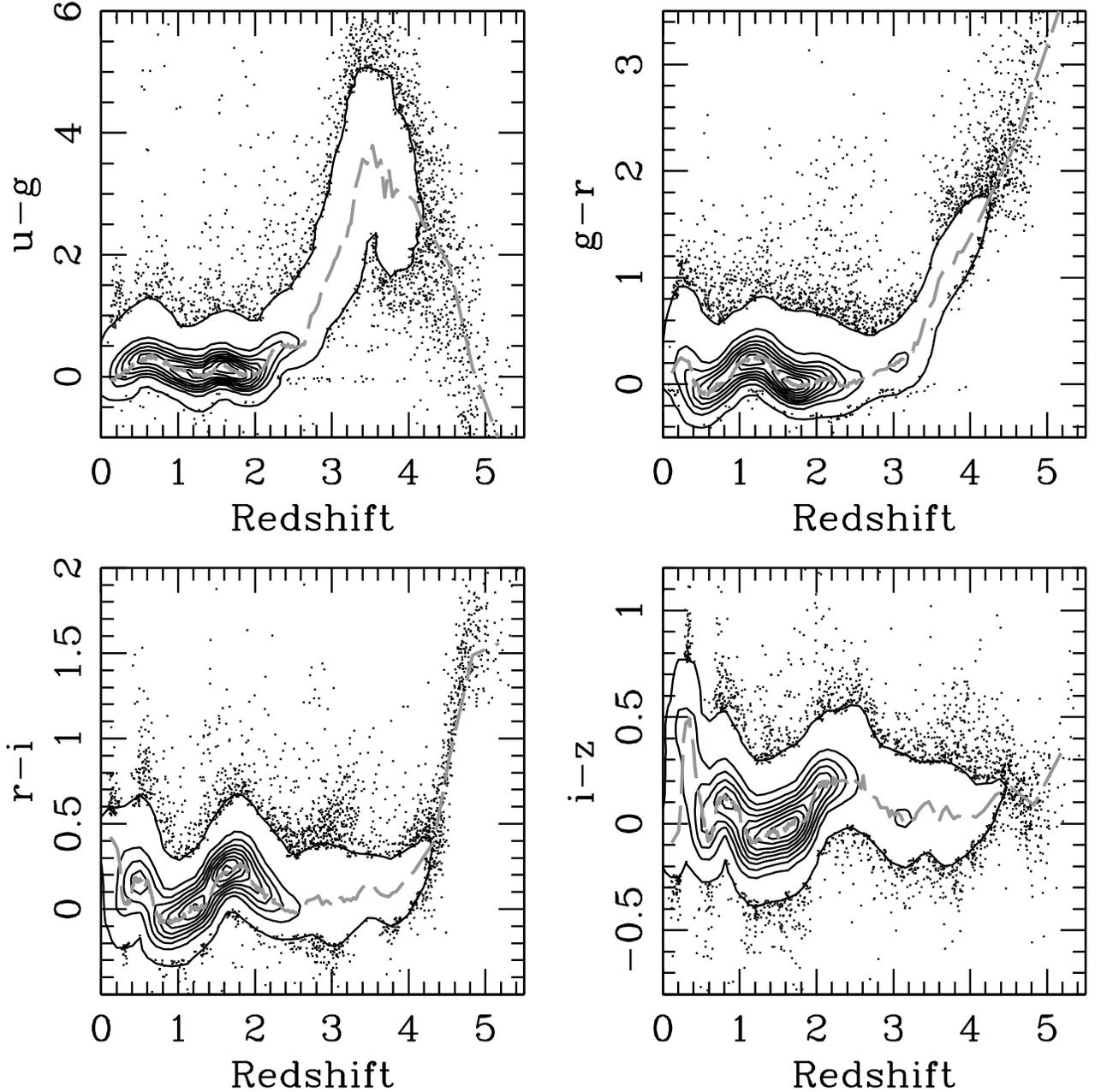}
\figcaption{
The quasar color-redshift relation for the DR5 quasars (photometry corrected
for Galactic extinction).  Contours are used to represent the distribution
when the density of points causes the points to overlap.  The panels
present the four standard SDSS colors; the dashed gray lines are the modal
relations presented in Table~5.  The influence of emission lines on the colors
is readily apparent (in particular H$\alpha$ in the $(i-z)$ panel).
The tightness of the correlations breaks down when the Lyman~$\alpha$ forest
region dominates the bluer of the two passbands (e.g., above redshifts
of 2.2 in the $(u-g)$ relation).
\label{Figure 7 }
}
\end{figure}

\begin{figure}
\includegraphics[angle=90,scale=0.70,viewport= 0 0 500 700]{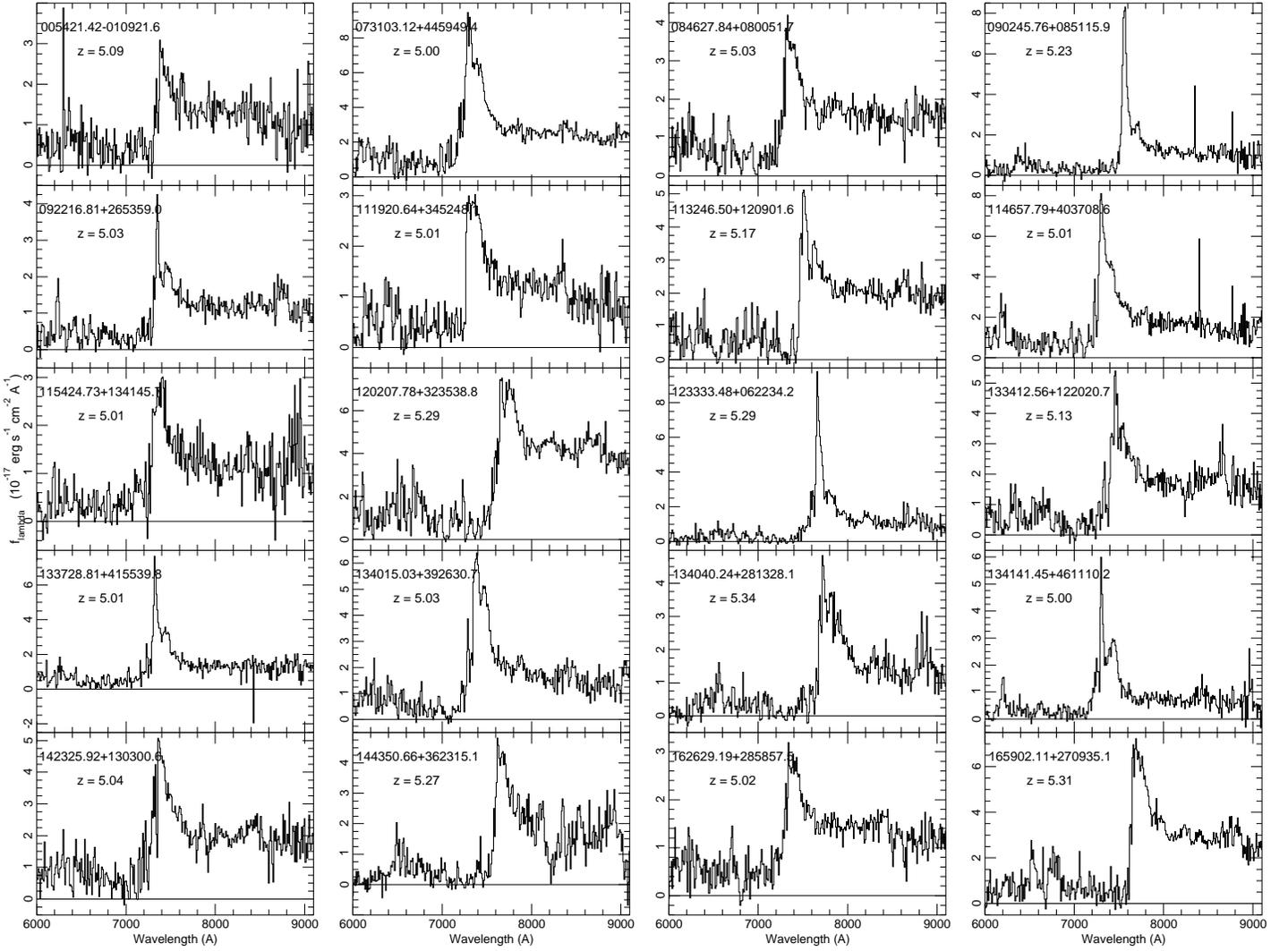}
\figcaption{
SDSS spectra of the 20 new quasars with the highest redshifts ($z \ge 4.99$).
The spectra have been rebinned to \hbox{10 \AA\ pixel$^{-1}$} for display
purposes.  The wavelength region below~6000~\AA\ has been removed because
of the lack of signal below rest frame wavelengths of 1000~\AA\ in
these objects.  Five of the quasars have redshifts larger than~5.25.
\label{Figure 8 }
}
\end{figure}

\begin{figure}
\includegraphics[angle=90,scale=0.70,viewport= 0 0 500 700]{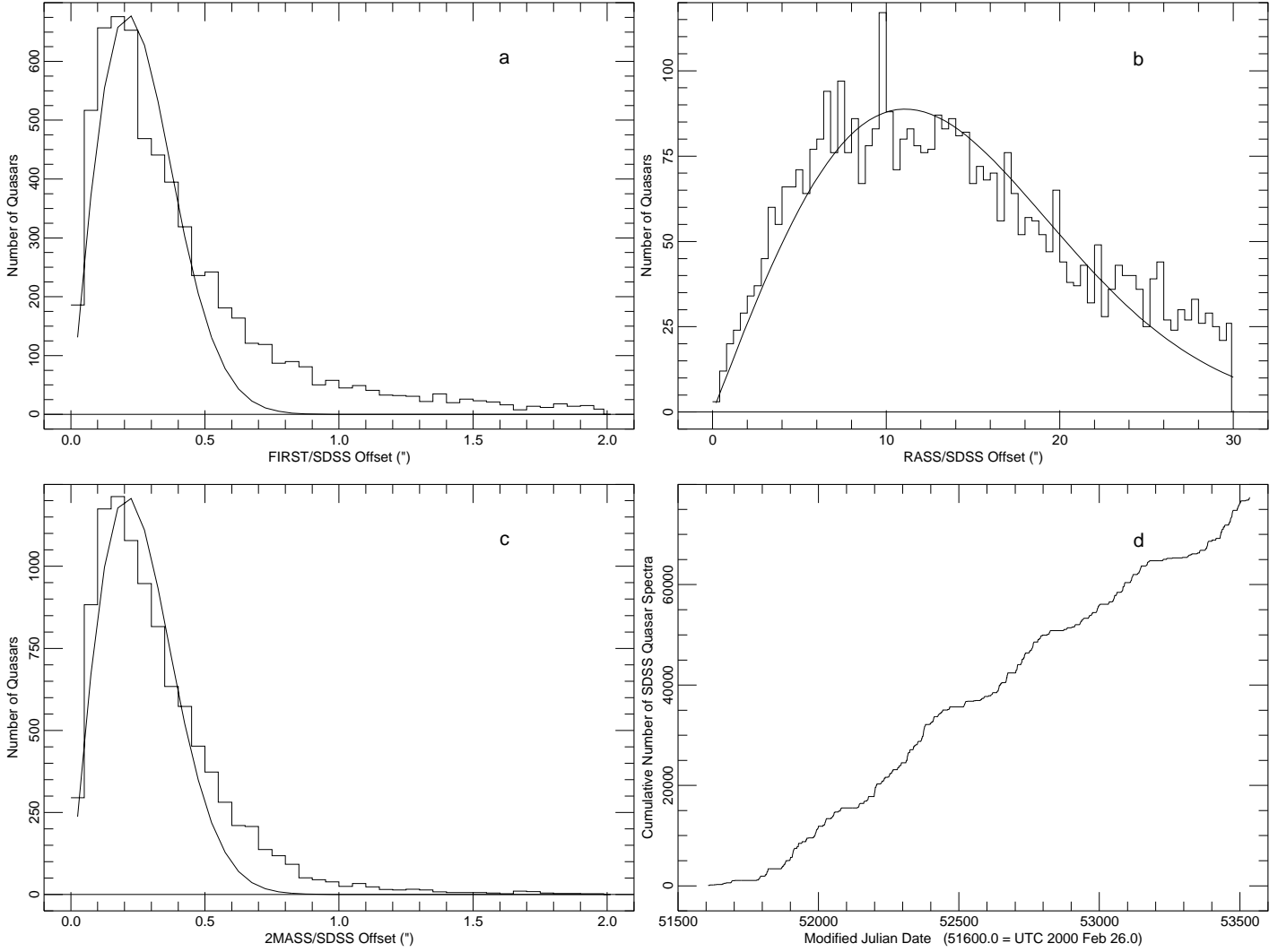}
\figcaption{
a)~Offsets between the~6226~SDSS and FIRST matches; the
matching radius was set to~2.0$''$.  The smooth curve is the expected
distribution for a set of matches if the offsets between the objects are
described by a Rayleigh distribution with \hbox{$\sigma = 0.21''$}
(best fit for points with separations of less than~1.0$''$).
b)~Offsets between the~4133 SDSS and RASS FSC/BSC matches; the
matching radius was set to~30$''$.  The smooth curve is the Rayleigh
distribution fit \hbox{($\sigma = 11.1''$)} to all of the points.
c)~Offsets between the~9824~SDSS and 2MASS matches; the
matching radius was set to~2$''$.  The smooth curve is a Rayleigh distribution
with $\sigma$~=~0.21$''$ based on the points
with separations smaller than~1.0$''$.
d)~The cumulative number of DR5 quasars as a function of time.  The horizontal
axis runs from February~2000 to June~2005.  The periodic structure
in the curve is caused by the yearly summer maintenance schedule.
The total number of objects in
the catalog \hbox{is 77,429.}
\label{Figure 9 }
}
\end{figure}

\clearpage


\begin{deluxetable}{rcl}
\tabletypesize{\small}
\tablewidth{0pt}
\tablecaption {SDSS DR5 Quasar Catalog Format}
\tablehead{
\colhead{Column} &
\colhead{Format} &
\colhead{Description}
}

\startdata
   1  &  A18  &
SDSS DR5 Designation   \ \ \ \ hhmmss.ss+ddmmss.s  \ \ \ (J2000) \\
   2  &  F11.6  &   Right Ascension in decimal degrees (J2000) \\
   3  &  F11.6  &   Declination in decimal degrees (J2000) \\
   4  &  F7.4 &   Redshift \\
   5  &   F7.3  &    BEST PSF $u$ magnitude
(not corrected for Galactic extinction) \\
   6  &   F6.3  &    Error in BEST PSF $u$ magnitude \\
   7  &   F7.3  &    BEST PSF $g$ magnitude
(not corrected for Galactic extinction) \\
   8  &   F6.3  &    Error in BEST PSF $g$ magnitude \\
   9  &   F7.3  &    BEST PSF $r$ magnitude
(not corrected for Galactic extinction) \\
  10  &   F6.3  &    Error in BEST PSF $r$ magnitude \\
  11  &   F7.3  &    BEST PSF $i$ magnitude
(not corrected for Galactic extinction) \\
  12  &   F6.3  &    Error in BEST PSF $i$ magnitude \\
  13  &   F7.3  &    BEST PSF $z$ magnitude
(not corrected for Galactic extinction) \\
  14  &   F6.3  &    Error in BEST PSF $z$ magnitude \\
  15  &   F7.3  &    Galactic extinction in $u$ band \\
  16  &   F7.3  &    $\log N_H$  (logarithm of Galactic H I column density) \\
  17  &   F7.3  &    FIRST peak flux density at 20 cm expressed as AB
magnitude; \\
& & \ \ \ \ \ 0.0 is no detection, $-1.0$ source is not in FIRST area \\
  18  &   F8.3  &    S/N of FIRST flux density \\
  19  &   F7.3  &    SDSS-FIRST separation in arc seconds \\
  20  &   I3    &   $> 3\sigma$ FIRST flux at optical position
but no FIRST counterpart within 2$''$ (0 or 1) \\
  21  &   I3    &   FIRST source located 2$''$-30$''$ from optical position
(0 or 1) \\
  22  &   F8.3  &   log RASS full band count rate; $-9.0$ is no detection \\
  23  &   F7.3  &   S/N of RASS count rate \\
  24  &   F7.3  &   SDSS-RASS separation in arc seconds \\
  25  &   F7.3  &   $J$ magnitude (2MASS);
0.0 indicates no 2MASS detection \\
  26  &   F6.3  &   Error in $J$ magnitude (2MASS) \\
  27  &   F7.3  &   $H$ magnitude (2MASS);
0.0 indicates no 2MASS detection \\
  28  &   F6.3  &   Error in $H$ magnitude (2MASS) \\
  29  &   F7.3  &   $K$ magnitude (2MASS);
0.0 indicates no 2MASS detection \\
  30  &   F6.3  &   Error in $K$ magnitude (2MASS) \\
  31  &   F7.3  &   SDSS-2MASS separation in arc seconds \\
  32  &   F8.3  &   $M_{i}$ ($H_0$ = 70 km s$^{-1}$ Mpc$^{-1}$,
$\Omega_M = 0.3$, $\Omega_{\Lambda} = 0.7$, $\alpha_{\nu} = -0.5$) \\
  33  &   F7.3  &   $\Delta(g-i) = (g-i) - \langle (g-i)
\rangle_{\rm redshift}$ (Galactic extinction corrected)  \\
  34  &   I3  &   Morphology flag \ \ \ 0 = point source \ \ \ 1 = extended \\
  35  &   I3  &   SDSS SCIENCEPRIMARY flag  (0 or 1) \\
  36  &   I3  &   SDSS MODE flag (blends, overlapping scans; 1, 2, or 3) \\
  37  &   I3  &   Selected with final quasar algorithm (0 or 1) \\
  38  &  I12  &   Target Selection Flag (BEST) \\
  39  &   I3  &   Low-$z$ Quasar selection flag (0 or 1) \\
  40  &   I3  &   High-$z$ Quasar selection flag (0 or 1) \\
  41  &   I3  &   FIRST selection flag (0 or 1) \\
  42  &   I3  &   {\it ROSAT} selection flag (0 or 1) \\
  43  &   I3  &   Serendipity selection flag (0 or 1) \\
  44  &   I3  &   Star selection flag (0 or 1) \\
  45  &   I3  &   Galaxy selection flag (0 or 1) \\
  46  &   I6  &   SDSS Imaging Run Number of photometric measurements \\
  47  &   I6  &   Modified Julian Date of imaging observation \\
  48  &   I6  &   Modified Julian Date of spectroscopic observation \\
  49  &   I5  &   Spectroscopic Plate Number \\
  50  &   I5  &   Spectroscopic Fiber Number \\
  51  &   I4  &   SDSS Photometric Processing Rerun Number \\
  52  &   I3  &   SDSS Camera Column Number \\
  53  &   I5  &   SDSS Field Number \\
  54  &   I5  &   SDSS Object Number \\
  55  &  I12  &   Target Selection Flag (TARGET) \\
  56  &   I3  &   Low-$z$ Quasar selection flag (0 or 1) \\
  57  &   I3  &   High-$z$ Quasar selection flag (0 or 1) \\
  58  &   I3  &   FIRST selection flag (0 or 1) \\
  59  &   I3  &   {\it ROSAT} selection flag (0 or 1) \\
  60  &   I3  &   Serendipity selection flag (0 or 1) \\
  61  &   I3  &   Star selection flag (0 or 1) \\
  62  &   I3  &   Galaxy selection flag (0 or 1) \\
  63  &   F7.3  &    TARGET PSF $u$ magnitude
(not corrected for Galactic extinction) \\
  64  &   F6.3  &    TARGET Error in PSF $u$ magnitude \\
  65  &   F7.3  &    TARGET PSF $g$ magnitude
(not corrected for Galactic extinction) \\
  66  &   F6.3  &    TARGET Error in PSF $g$ magnitude \\
  67  &   F7.3  &    TARGET PSF $r$ magnitude
(not corrected for Galactic extinction) \\
  68  &   F6.3  &    TARGET Error in PSF $r$ magnitude \\
  69  &   F7.3  &    TARGET PSF $i$ magnitude
(not corrected for Galactic extinction) \\
  70  &   F6.3  &    TARGET Error in PSF $i$ magnitude \\
  71  &   F7.3  &    TARGET PSF $z$ magnitude
(not corrected for Galactic extinction) \\
  72  &   F6.3  &    TARGET Error in PSF $z$ magnitude \\
  73  &  I21    &    Spectroscopic Identification flag (64-bit integer) \\
  74  &   1X, A25 &   Object Name for previously known quasars \\
 & & \ \ \ ``SDSS" designates previously published SDSS object \\
\enddata

\end{deluxetable}

\clearpage

\begin{deluxetable}{crrrcrrrrrrrrr}
\tabletypesize{\small}
\rotate
\tablewidth{0pt}
\tablecaption {The SDSS Quasar Catalog IV$^{a}$}
\tablehead{
\colhead{Object (SDSS J)} &
\colhead{R.A. (deg)} &
\colhead{Dec (deg)} &
\colhead{Redshift} &
\multicolumn{2}{c}{$u$} &
\multicolumn{2}{c}{$g$} &
\multicolumn{2}{c}{$r$} &
\multicolumn{2}{c}{$i$} &
\multicolumn{2}{c}{$z$}
}

\startdata
000006.53+003055.2 & 0.027228 &   0.515349 & 1.8227 & 20.389 & 0.066 &
20.468 & 0.034 & 20.332 & 0.037 & 20.099 & 0.041 & 20.053 & 0.121 \\
000008.13+001634.6 & 0.033898 &   0.276304 & 1.8365 & 20.233 & 0.054 &
20.200 & 0.024 & 19.945 & 0.032 & 19.491 & 0.032 & 19.191 & 0.068 \\
000009.26+151754.5 & 0.038605 &  15.298476 & 1.1986 & 19.921 & 0.042 &
19.811 & 0.036 & 19.386 & 0.017 & 19.165 & 0.023 & 19.323 & 0.069 \\
000009.38+135618.4 & 0.039088 &  13.938447 & 2.2400 & 19.218 & 0.026 &
18.893 & 0.022 & 18.445 & 0.018 & 18.331 & 0.024 & 18.110 & 0.033 \\
000009.42$-$102751.9 & 0.039269 & $-$10.464428 & 1.8442 & 19.249 & 0.036 &
19.029 & 0.027 & 18.980 & 0.021 & 18.791 & 0.018 & 18.751 & 0.047 \\
\enddata

\tablenotetext{a}{
Table 2 is presented in its entirety in the electronic edition of the
Astronomical Journal.  A portion is shown here for guidance regarding
its form and content.  The full catalog
contains 74 columns of information
on 77,429 quasars.}

\end{deluxetable}

\clearpage

\begin{deluxetable}{lrrrr}
\tablewidth{0pt}
\tablecaption {Spectroscopic Target Selection}
\tablehead{
\colhead{} &
\colhead{TARGET} &
\colhead{TARGET} &
\colhead{BEST} &
\colhead{BEST} \\
\colhead{} &
\colhead{} &
\colhead{Sole} &
\colhead{} &
\colhead{Sole} \\
\colhead{Class} &
\colhead{Selected} &
\colhead{Selection} &
\colhead{Selected} &
\colhead{Selection}
}

\startdata
Low-$z$      &  49010  &  16422   &    46460 &   14444 \\
High-$z$     &  16383  &   5327   &    16757 &    4411 \\
FIRST        &   3501  &    226   &     3619 &     209 \\
{\it ROSAT}  &   4817  &    380   &     4918 &     492 \\
Serendipity  &  42109  &  15729   &    41042 &   15950 \\
Star         &   1970  &    187   &      820 &     162 \\
Galaxy       &    536  &     99   &      601 &      80 \\
\enddata

\end{deluxetable}

\clearpage

\begin{deluxetable}{lrlr}
\tablewidth{0pt}
\tablecaption {
Quasars with $\vert z_{\rm DR5} - z_{\rm DR3} \vert > 0.1$}
\tablehead{
\colhead{SDSS J} &
\colhead{$z_{\rm DR5}$} &
\colhead{SDSS J} &
\colhead{$z_{\rm DR5}$}
}

\startdata
005508.55$-$105206.2 &  1.381 & 133028.12+600811.7 &  1.992 \\
013413.55+142900.1 &  1.195 & 133951.94+481651.3 &  0.911 \\
031712.23$-$075850.3 &  2.696 & 134048.37+433359.8 &  2.069 \\
075052.59+300334.1 &  3.990 & 135833.05+634122.6 &  3.180 \\
075132.75+350535.0 &  2.077 & 140012.65+595823.3 &  2.061 \\
\noalign{\smallskip}
083503.79+322242.0 &  0.728 & 140223.63+463604.9 &  0.925 \\
085339.64+372203.6 &  1.950 & 140327.91+613654.2 &  2.023 \\
090902.73+355334.8 &  1.638 & 141230.28+471103.7 &  2.078 \\
091025.25+365921.3 &  2.004 & 142010.28+604722.3 &  1.345 \\
092415.87+424632.2 &  0.559 & 143702.47+613437.0 &  2.064 \\
\noalign{\smallskip}
093557.85+005528.1 &  1.301 & 144939.30+534212.1 &  1.805 \\
093935.08$-$000801.1 &  0.909 & 151307.26$-$000559.3 &  2.030 \\
094326.48+460226.8 &  2.093 & 151422.99+481936.3 &  2.071 \\
100415.17+415802.6 &  1.977 & 153257.67+422047.1 &  1.950 \\
102117.71+623010.1 &  1.949 & 160320.97+315248.3 &  0.727 \\
\noalign{\smallskip}
103039.95+510923.3 &  1.649 & 165806.76+611858.9 &  2.631 \\
103219.66+563456.8 &  2.017 & 170929.58+323826.9 &  1.902 \\
115917.62+100921.5 &  2.028 & 205058.45+004709.9 &  0.932 \\
124345.10+492645.3 &  1.982 & 212744.12+005720.3 &  4.386 \\
131810.57+585416.9 &  1.900 & 225246.43+142525.8 &  4.904 \\
\enddata

\end{deluxetable}

\clearpage

\begin{deluxetable}{rrrrrrrr}
\tablewidth{0pt}
\tablecaption {Quasar Colors as a Function of Redshift$^{a}$}
\tablehead{
\colhead{$z_{\rm bin}$} &
\colhead{$\langle z \rangle $} &
\colhead{$N_{\rm QSO}$} &
\colhead{$(g-i)$} &
\colhead{$(u-g)$} &
\colhead{$(g-r)$} &
\colhead{$(r-i)$} &
\colhead{$(i-z)$}
}

\startdata
 0.18 & 0.181 & 183 &  0.567 & $-$0.065 &  0.197 &  0.379 & $-$0.037 \\
 0.21 & 0.210 & 290 &  0.580 &  0.032 &  0.223 &  0.355 & $-$0.034 \\
 0.24 & 0.240 & 394 &  0.513 &  0.000 &  0.236 &  0.267 &  0.115 \\
 0.27 & 0.270 & 406 &  0.289 &  0.055 &  0.231 &  0.077 &  0.397 \\
 0.30 & 0.301 & 484 &  0.236 &  0.067 &  0.219 &  0.033 &  0.472 \\
\enddata

\tablenotetext{a}{
Table 5 is presented in its entirety in the electronic edition of the
Astronomical Journal.  A portion is shown here for guidance regarding
its form and content.}

\end{deluxetable}

\clearpage

\begin{deluxetable}{ccccr}
\tablewidth{0pt}
\tablecaption {Candidate Binary Quasars}
\tablehead{
\colhead{Quasar 1} &
\colhead{Quasar 2} &
\colhead{$z_1$} &
\colhead{$z_2$} &
\colhead{$\Delta \theta$ $''$}
}

\startdata
001201.87+005259.7 & 001202.35+005314.0 & 1.652 & 1.642 &  16.0 \\
011757.99+002104.1 & 011758.83+002021.4 & 0.612 & 0.613 &  44.5 \\
014110.40+003107.1 & 014111.62+003145.9 & 1.879 & 1.882 &  42.9 \\
024511.93$-$011317.5 & 024512.12$-$011313.9 & 2.463 & 2.460 &   4.5 \\
025813.65$-$000326.4 & 025815.54$-$000334.2 & 1.316 & 1.321 &  29.4 \\
025959.68+004813.6 & 030000.57+004828.0 & 0.892 & 0.900 &  19.6 \\
074336.85+205512.0 & 074337.28+205437.1 & 1.570 & 1.565 &  35.5 \\
074759.02+431805.4 & 074759.66+431811.5 & 0.501 & 0.501 &   9.2 \\
082439.83+235720.3 & 082440.61+235709.9 & 0.536 & 0.536 &  14.9 \\
085625.63+511137.0 & 085626.71+511117.8 & 0.543 & 0.543 &  21.8 \\
090923.12+000203.9 & 090924.01+000211.0 & 1.884 & 1.865 &  15.0 \\
095556.37+061642.4 & 095559.02+061701.8 & 1.278 & 1.273 &  44.0 \\
110357.71+031808.2 & 110401.48+031817.5 & 1.941 & 1.923 &  57.3 \\
111610.68+411814.4 & 111611.73+411821.5 & 2.980 & 2.971 &  13.8 \\
113457.73+084935.2 & 113459.37+084923.2 & 1.533 & 1.525 &  27.1 \\
121840.47+501543.4 & 121841.00+501535.8 & 1.457 & 1.455 &   9.1 \\
165501.31+260517.5 & 165502.02+260516.5 & 1.881 & 1.892 &   9.6 \\
215727.26+001558.4 & 215728.35+001545.5 & 2.540 & 2.553 &  20.8 \\
\enddata

\tablenotetext{a}{The quasar pairs were selected by a redshift difference of
less than 0.02 and an angular separation less than 60$''$.}

\end{deluxetable}

\clearpage
\end{document}